\documentclass[letterpaper,twocolumn,10pt]{article}
\usepackage{usenix-2020-09}

\usepackage{prp-macros}

\usepackage{pdfpages}

\usepackage[normalem]{ulem}

\usepackage[font={small, bf, sf}, textfont={small, sf}]{caption}
\usepackage{float}

\usepackage[small, compact]{titlesec}

\usepackage[available,functional,reproduced]{usenixbadges}

\usepackage{tikz}
\usetikzlibrary{external}
\usetikzlibrary{positioning}
\usepackage{pgfplots}
\usetikzlibrary{calc}
\usepackage{pifont}
\usepgfplotslibrary{fillbetween}
\usetikzlibrary{shapes.arrows}
\usetikzlibrary{arrows.meta}
\usetikzlibrary{arrows, backgrounds} 
\usetikzlibrary{intersections,through, decorations}
\usetikzlibrary{chains,patterns,shadows,fit,backgrounds}

\definecolor[named]{ACMBlue}{cmyk}{1,0.1,0,0.1}
\definecolor[named]{ACMYellow}{cmyk}{0,0.16,1,0}
\definecolor[named]{ACMOrange}{cmyk}{0,0.42,1,0.01}
\definecolor[named]{ACMRed}{cmyk}{0,0.90,0.86,0}
\definecolor[named]{ACMLightBlue}{cmyk}{0.49,0.01,0,0}
\definecolor[named]{ACMGreen}{cmyk}{0.20,0,1,0.19}
\definecolor[named]{ACMPurple}{cmyk}{0.55,1,0,0.15}
\definecolor[named]{ACMDarkBlue}{cmyk}{1,0.58,0,0.21}
\colorlet{Yellow}{ACMYellow}
\colorlet{Green}{ACMGreen}
\colorlet{Red}{ACMRed}
\colorlet{SkyBlue}{ACMLightBlue}
\definecolor[named]{Gray}{cmyk}{0,0,0,0.25}
\definecolor[named]{Grey}{cmyk}{0,0,0,0.25} 

\makeatletter
\tikzset{
        hatch distance/.store in=\hatchdistance,
        hatch distance=5pt,
        hatch thickness/.store in=\hatchthickness,
        hatch thickness=5pt
        }
\pgfdeclarepatternformonly[\hatchdistance,\hatchthickness]{north east hatch}
    {\pgfqpoint{-1pt}{-1pt}}
    {\pgfqpoint{\hatchdistance}{\hatchdistance}}
    {\pgfpoint{\hatchdistance-1pt}{\hatchdistance-1pt}}%
    {
        \pgfsetcolor{\tikz@pattern@color}
        \pgfsetlinewidth{\hatchthickness}
        \pgfpathmoveto{\pgfqpoint{0pt}{0pt}}
        \pgfpathlineto{\pgfqpoint{\hatchdistance}{\hatchdistance}}
        \pgfusepath{stroke}
    }
\makeatother

\tikzset{
    *|/.style={
        to path={
            (perpendicular cs: horizontal line through={(\tikztostart)},
                                 vertical line through={(\tikztotarget)})
            -- (\tikztotarget) \tikztonodes
        }
    },
    ^|/.style={
        to path={
            (perpendicular cs: vertical line through={(\tikztostart)},
                                 horizontal line through={(\tikztotarget)})
            -- (\tikztotarget) \tikztonodes
        }
    },
    vecArrow/.style={
		thick,
  		decoration={markings,mark=at position 1 with {\arrow[scale=2,thin]{open triangle 90}}},
 		double distance=3mm, shorten >= 2.6mm,
  		preaction = {decorate},
  		postaction = {draw,line width=3.0mm, white,shorten >= 2.5mm}
  	},
	innerWhite/.style={
  		semithick, 
  		white,
  		line width=3mm, 
  		shorten >= 2.5mm
  	}
}

\newcommand{\coffer}{\change{cVM}\xspace}
\newcommand{\monitor}{\change{Intravisor}\xspace}
\newcommand{\coffers}{\change{cVMs}\xspace}
\newcommand{\magic}{\coffer}

\newcommand{\todo}[1]{\textbf{\color{red}[\textsc{todo:} #1]}}

\usepackage[shortcuts,acronyms, nohypertypes={acronym,notation}]{glossaries-extra}

\setabbreviationstyle[acronym]{long-short}
\newacronym{file}{CP\_File}{CAP\_File}
\newacronym{call}{CP\_Call}{CAP\_Call}
\newacronym{stream}{CP\_Stream}{CAP\_Stream}

\newacronym{pcc}{PCC}{Program Counter Capability}
\newacronym{ddc}{DDC}{Default Data Capability}
\newacronym{crpc}{CRPC}{Capability Remote Function Call}

\usepackage{tikz}
\usetikzlibrary{positioning}
\usepackage{pgfplots}
\usetikzlibrary{calc}
\usepackage{xspace}
\usepgfplotslibrary{fillbetween}
\usetikzlibrary{shapes.arrows}
\usetikzlibrary{arrows.meta}
\usetikzlibrary{arrows, backgrounds} 
\usetikzlibrary{intersections,through, decorations}
\usetikzlibrary{chains,patterns,shadows,fit,backgrounds}
\usepackage{tikz-uml}

\usepackage{cleveref}

\begin{document}

\title{CAP-VMs: Capability-Based Isolation and Sharing in the Cloud}


\author{
{\rm Vasily\ A.\ Sartakov}\\
Imperial College London
\and
{\rm Llu\'is\ Vilanova}\\
Imperial College London
 \and
 {\rm David Eyers}\\
University of Otago
 \and
 {\rm Takahiro Shinagawa}\\
The University of Tokyo
 \and
 {\rm  Peter Pietzuch}\\
Imperial College London
} 

\date{}

\maketitle
\pagestyle{empty}

\thispagestyle{empty}

\begin{abstract}


Cloud stacks must isolate application components, while permitting efficient data sharing between components deployed on the same physical host. Traditionally, the MMU enforces isolation and permits sharing at page granularity. MMU approaches, however, lead to cloud stacks with large TCBs in kernel space, and page granularity requires inefficient OS interfaces for data sharing. Forthcoming CPUs with hardware support for \emph{memory capabilities} offer new opportunities to implement isolation and sharing at a finer granularity.

We describe \emph{\coffers}, a new \change{VM-like} abstraction that uses memory capabilities to isolate application components while supporting efficient data sharing, all without mandating application code to be capability-aware. \coffers share a single virtual address space safely, each having only capabilities to access its own memory. A \coffer may include a library OS, thus minimizing its dependency on the cloud environment. \coffers efficiently exchange data through two capability-based primitives assisted by a small trusted monitor: (i)~an asynchronous read\slash write interface to buffers shared between \coffers; and (ii)~a call interface to transfer control between \coffers. Using these two primitives, we build more expressive mechanisms for efficient cross-\coffer communication. Our prototype implementation using CHERI RISC-V capabilities shows that \coffers isolate services (Redis and Python) with low overhead while improving data sharing.

\end{abstract}


\section{Introduction}
\label{sec:intro}




\change{Cloud environments require application compartmentalization. Today, isolation between application components is enforced by virtual machines~(VMs)~\cite{xen, kvm, esxi} and containers~\cite{lxc, docker}, either separately or in combination. Yet, current applications push the limits of these mechanisms in terms of performance and security: when application components communicate heavily with each other, VMs and containers add substantial overheads, even when they are co-located to improve communication performance; furthermore, the implementation of the isolation mechanisms may also rely on a large trusted computing base~(TCB).}



\emph{VMs} provide strong isolation through a relatively narrow hardware interface. Since a guest VM has its own OS kernel, its TCB can be reduced to a relatively small hypervisor, which multiplexes VM access to the hardware~\cite{nova}. Efficient inter-VM data sharing, however, is challenging to achieve due to performance and page granularity trade-offs~\cite{efficient_inter_VM_channels, Xen_event_channels}.


In contrast, \emph{containers} isolate processes into groups~\cite{lxc} and provide faster inter-process communication~(IPC) primitives, including pipes, shared memory, and sockets. Similar to VMs, they face problems of page-level sharing granularity and overheads due to frequent user/kernel transitions. Their richer IPC primitives for data sharing come at the cost of a larger TCB---a shared OS kernel implements both namespace isolation between process groups and complex IPC primitives, increasing the likelihood of security vulnerabilities.

Existing cloud stacks thus face a fundamental tension when application components are compartmentalized but must communicate. They must either copy data or modify page tables, both of which are expensive operations that involve a privileged intermediary, \eg a hypervisor or OS kernel, and lead to coarse-grained interfaces designed around page granularity.

\change{In this work, we explore a different approach to designing a cloud stack that isolates application components, while supporting efficient sharing. We ask the question ``if the hardware supported dynamic, low-overhead sharing of arbitrary-sized memory regions between otherwise isolated regions, how would this impact the cloud stack design?'' We exploit hardware support for \emph{memory capabilities}~\cite{10.1145/1353535.1346295, cheri}, which impose flexible bounds on all memory accesses, allowing components to be isolated without page table modifications or adherence to page boundaries. This offers a new opportunity to design memory sharing primitives between isolated compartments with zero-copy semantics.}

We describe \textbf{CAP-VMs (\coffers)}, a new \change{VM-like} abstraction for executing isolated components and sharing data across them. \coffers are enforced by a small TCB that uses memory capabilities to isolate and share data between compartments efficiently. Through the use of a \emph{hybrid} capability model~\cite{watson2015cheriComp}, \coffers avoid having to port application components to use capability instructions, circumventing compatibility issues that typically plague capability architectures.


Using memory capabilities as part of a cloud stack, however, raises new challenges: the cloud stack must (i)~support existing capability-unaware software without cumbersome code changes, bespoke compiler support, or manual management of capabilities across isolation boundaries; (ii)~remain compatible with existing OS abstractions, \eg POSIX interfaces, all while keeping the TCB small; and (iii)~offer efficient IPC-like primitives for otherwise untrusted components to share data safely and take advantage of the potential zero-copy sharing enabled by capabilities.


To address the above challenges, \coffers make the following design contributions:


\mypar{(1)~Strong isolation through capabilities} Multiple \coffers share a single virtual address space safely through capabilities. Each \coffer is sandboxed by a pair of \emph{default} capabilities, which confine the accesses of all instructions inside a \coffer to its own memory boundaries. To avoid having to port existing application components to a capability architecture, \coffers allow them to execute unmodified by using CHERI's \emph{hybrid} capability architecture~\cite{watson2015cheriComp}, which integrates capabilities with a conventional MMU architecture. In addition, \coffers strictly limit how CHERI capabilities can be used to avoid known capability revocation overheads: \coffers are not permitted to store or export capabilities, and the transitions of communication capabilities \change[is]{are} controlled by a trusted component. 


 
\mypar{(2)~Bespoke OS support through a library OS} \coffers are self-contained with a small TCB, reducing reliance on the external cloud stack, while providing POSIX compatibility. They include a bespoke \emph{library OS} with POSIX interfaces for, \eg filesystem and network operations with cryptography for transparent protection, which is protected from application code using capabilities. In the library OS, each \coffer implements its own namespace for filesystem objects, virtual devices, cryptographic I/O keys etc. Only low-level resources, \eg execution contexts for threads and I/O device operations, are shared and provided by an external host OS kernel.


\mypar{(3)~Efficient data sharing primitives} \coffers offer two low-level primitives to share data efficiently without exposing application code to capabilities, which are hidden behind a small, trusted \emph{\monitor}: (i)~a \emph{\acs{file}} API allows application components to share arbitrary buffers through an asynchronous read/write interface. Under the hood, the \coffer implementation uses capability-aware instructions to exchange the rights to safely access each other's memory, and read/write data at byte granularity at the cost of a single memory copy (whereas traditional file-oriented IPC would require two copies); and (ii)~a \emph{\acs{call}} API transfers control between \coffers, which, \eg can be used to implement synchronization mechanisms. By combining these two primitives, higher-level APIs are possible: (iii)~a \emph{\acs{stream}} API supports efficient stream-oriented data exchange between \coffers with one memory copy.


\tinyskip

\noindent
We implement \coffers on the CHERI RISC-V64 architecture, executable on FPGA hardware with CHERI support and multi-core RISC-V hardware. Our evaluation shows that \coffers provide a practical isolation abstraction with efficient data sharing: using the \acs{stream} API for inter-\coffer communication reduces latency for Redis by up to 54\% compared to classical socket interfaces, and reduces its standard deviation by up to 2.1$\times$. When isolating a cryptography component of a Python-based service, \coffers introduce an overhead of up to 12\% compared to a monolithic baseline.

\section{Hardware Isolation Support}
\label{sec:background}

Next we survey the design space for isolation and sharing in cloud environments in more detail~(\S\ref{sec:background:isolation_sharing}), provide background on capability support on modern hardware~(\S\ref{sec:background:cheri}), and describe our threat model~(\S\ref{sec:background:threat_model}).

\subsection{Isolation and sharing in the cloud}
\label{sec:background:isolation_sharing}




We argue that VMs and containers are two extremes of component isolation. VMs virtualize hardware interfaces such as page tables, instructions, traps, and physical device interfaces to manage both isolation and communication; containers virtualize pure software interfaces such as processes, files, and sockets for the same purposes.

\mypar{Compatibility} Both VMs and containers are compatible with existing applications, which is critical for adoption in cloud environments. VMs can execute an unmodified guest OS on top of a hypervisor, making virtualization transparent to applications inside VMs. Conversely, containers execute unmodified applications on top of the same host OS kernel that manages other containerized and non-containerized applications. In both cases, OS interfaces and semantics used by the virtualized applications remain unmodified compared to a non-virtualized environment.


But the compatibility offered by these technologies lowers communication performance, which is often exacerbated as we try to achieve better isolation between components.

\mypar{Isolation} Despite strict isolation between the memory of containers, there is a lack of isolation of the TCB that manages the virtualization mechanism itself. Conventional container platforms, \eg Linux containers~\cite{lxc}, share privileged state, as they employ namespace virtualization: the OS kernel creates separate process identifiers, devices, filesystem views etc., which offer the illusion that a process group exists in isolation. In reality, containers share kernel data structures, and privilege escalation inside one container may lead to the compromise of all containers~\cite{CVE-2021-21284, CVE-2013-6441}. In comparison, VMs are virtualized through narrower interfaces, resulting in a conceptually simpler hypervisor that is harder to compromise~\cite{nova, muen}.

Unfortunately, stronger isolation comes at a performance price from both known hardware inefficiencies~\cite{MOTIKA201236, buell2013methodology, 10.1145/3307650.3322261} as well as less flexible mechanisms for data sharing.


\mypar{Sharing} \change{Components of cloud applications typically use networking as a means of communication. Even if multiple components are co-located on the same host, they may use a reliable network transport protocol, \eg TCP\@. While this helps with scalability, it adds overhead for co-located components, making optimizations based on direct memory sharing attractive.} Both VMs and containers use page-based memory isolation, which limits the performance of memory sharing: mechanisms must be aware of page boundaries to avoid leaking sensitive data, and page table modifications for on-demand sharing are known to be expensive~\cite{villavieja11:_didi}.

\change{Co-location opens up two avenues for performance improvements: (1)~sharing can transparently speed up communication of co-located components~\cite{ren16:_shmem_opt_vm, ning13:_virt_io_opt}; and (2)~new communication interfaces can be tailored toward efficient sharing between components.}

\subsection{CHERI capability architecture}
\label{sec:background:cheri}



\change{In cloud applications with many services~\cite{deathstarbench}, traditional network-based communication shows its performance limits between tightly-coupled components~\cite{kogias19:_r2p2}. Therefore, we aim to co-locate components and design a cloud stack with efficient isolation and communication interfaces and mechanisms. This requires, however, new hardware support for isolation and sharing that is free of the ``MMU tax'' of page-level privileged memory protection.}

\change{\emph{Memory capabilities}~\cite{memory_caps} are a protection and sharing mechanism supported by the hardware. The \emph{CHERI} architecture~\cite{cheri, cheriIsa} implements capabilities as an alternative to traditional memory pointers. A capability is stored in memory or registers, and encodes an address range with permissions, \eg referring to a read-only buffer or a callable function.}

CHERI protects capabilities by enforcing three properties: (1)~\emph{provenance validity} ensures that a capability can only be ``derived'', \ie constructed, from another valid capability, \ie \change{it is not possible to cast an arbitrary byte sequence to a capability}; (2)~\emph{capability integrity} means that capabilities stored in memory cannot be modified, \change{which CHERI achieves through transparent memory tagging~\cite{cheri}}; and (3)~\emph{capability monotonicity} requires that, if a capability is stored in a register, its bounds and permissions can only be reduced, \eg a read-only capability cannot be turned into a read-write one.


\mypar{Building capability-based compartments} CHERI capabilities can be used to compartmentalize software components, \eg plugins or libraries in a program, by giving each capabilities to separate memory regions. The above properties enforced by CHERI ensure that compartments can coexist in the same address space, and remain isolated as long as their initial capabilities point to disjoint data and code in memory. The application can, of course, grant each compartment extra capabilities, \eg to allow particular cross-compartment memory accesses or function calls.

\mypar{Pure- and hybrid-cap code} CHERI distinguishes between two execution modes~\cite{watson2015cheriComp}: (i)~in \emph{pure-cap} mode, all pointers must be capabilities,\footnote{CHERI has separate registers for regular data and capabilities.} and code must use a new set of capability-aware instructions; and (ii)~in \emph{hybrid-cap} mode, code can mix ordinary and capability-aware instructions, which allows the coexistence of capability-unaware and pure-cap code via wrapping functions. This facilitates the incremental adoption of capabilities in software.

\change{When accessing memory, pure-cap code must use new instructions that use capability registers instead of regular registers. In addition, secure calls across capability-isolated components must use a \texttt{CInvoke} instruction, which requires a pair of capabilities: the target function address, and an arbitrary value that is meaningful to the callee function (\eg{} an identifier for an object managed by the callee).}

\change{To ensure that both capabilities are used correctly by \texttt{CInvoke}, \eg thwarting a malicious caller from passing a callee object identifier that was meant for a different callee function, the callee can ``seal'' pairs of capabilities together using the \texttt{CSeal} instruction. \texttt{CInvoke} only accepts correctly sealed pairs of capabilities.}


Hybrid-cap code relies on two new capability registers, the \emph{default data capability}~(\texttt{ddc}) and the \emph{program counter capability}~(\texttt{pcc}), which are used implicitly by capability-unaware instructions. The OS starts all processes by setting  \texttt{ddc} and \texttt{pcc} to the entire virtual address space. Capability-aware code then creates new capabilities from these registers, preserving CHERI's provenance, integrity and monotonicity properties.

Pure-cap code thus introduces compatibility challenges:

\begin{myitemize}
\item \change{All pointers in pure-cap code are capabilities that occupy 16\unit{bytes} instead of the ordinary 8\unit{bytes}, and must be 16-byte aligned. This decreases CPU cache effectiveness, and may require extra effort to align capability and non-capability elements in data structures.}


\item \change{It is not possible to cast between addresses and various types of capability-based pointers, because CHERI distinguishes between them and imposes bounds on pointers~\cite{watson2020cheri}. C/C++ code that uses raw casts---a commonly found idiom in low-level system software---requires substantial modifications. For example, the strict bounds in capabilities are typically incompatible with memory allocators that place metadata before allocated data.}

\item \change{While CHERI compresses capabilities, they can still result in memory bloat, because larger sizes are subject to coarser address discretization. Large allocations with capabilities may require stronger alignment and extra padding~\cite{woodruff2019cheri}.}

\item CHERI advocates for a trusted, system-wide \emph{garbage collector} to manage capabilities to dynamically-allocated memory~\cite{watson2015cheriComp}. \change{It is important to ensure that allocations are not reused while valid capabilities pointing to them still exist. Since new capabilities can be derived from existing ones, and stored on the heap, stack, and in registers, all capabilities derived from an allocation must be either invalidated (\ie revoked), or allocations cannot be reused while such capabilities are valid. A garbage collector (as opposed to expensive hardware support for capability revocation) addresses this issue, but it is a disruptive change in cloud environments, potentially leading to delays in resource reclamation and increased tail latencies.}
\end{myitemize}

Removing the need to use capability-aware code is important in cloud environments with limited control over tenant code. Therefore, we want to explore a design for a cloud stack that compartmentalizes application components using CHERI's hybrid-cap mode, without the disadvantages of pure capability-aware code.



\subsection{Threat model}
\label{sec:background:threat_model}

%

Cloud environments support multiple, isolated \change{application components}, and thus we consider attacks in which an attacker controls a malicious \change{component} that interferes with another \change{component} by probing interfaces or trying to escape its sandbox. We assume that the attacker has full control over the \change{application components and a library OS}, \eg by exploiting vulnerabilities inside the compartment or by executing arbitrary code that includes capability-aware instructions.


Our TCB includes the underlying host OS kernel, but the entire application stack (program, libraries and library OS) is considered untrusted. We assume that the CHERI hardware implementation is correct. We do not analyse side-channel attacks against CHERI, which is an important, yet orthogonal consideration that affects both the architectural and micro-architectural levels~\cite{watson2018capability}.

\begin{figure}[tb]
  \centering
  \resizebox{.99\linewidth}{!}{%
	\begin{tikzpicture}[->,>=stealth',shorten >=0pt,auto,node distance=1.5cm, thick,main node/.style={rectangle,draw, font=\normalsize,minimum size=5mm}]

	\node[main node, text width=2.5cm,align=center, minimum size=0.5cm,fill=ACMPurple!20] (SC10)  {
	\textbf{syscall interface}
	};

	\node[main node, text width=3.6cm,align=center, minimum size=0.5cm,draw=none] (SC1)  {
	};

	\node[main node, text width=2.5cm, align=center, minimum size=0.5cm,,fill=ACMPurple!20] [right = 2.0cm of SC1] (SC2) {\textbf{syscall interface}};


	\node[main node, text width=3.6cm, align=center, minimum size=1.0cm,fill=ACMPurple!20] [below = 0.25cm of SC1] (L1) {\textbf{Library OS} \\(namespace+environment)};

	\node[main node, text width=3.6cm, align=center, minimum size=1.0cm,fill=ACMPurple!20] [below = 0.25cm of SC2] (L2) {\hspace{0.75cm}\textbf{Library OS}};
	\node[main node, text width=2.75cm, align=center, minimum size=0.5cm,fill=ACMLightBlue!40] [below = 0.25cm of L2] (H2) {\textbf{hostcall interface}};

	\node[main node, text width=2.75cm, align=center, minimum size=0.5cm,fill=ACMLightBlue!40] [below = 0.25cm of L1] (H1) {\textbf{hostcall interface}};
	\node[main node, text width=9.5cm, align=center, minimum size=0.5cm,fill=ACMLightBlue!40] at ([yshift=-0.75cm]$(H1)!0.5!(H2)$) (M) {\textbf{\monitor}};

	\node[main node, text width=9.5cm, align=center, minimum size=0.5cm, fill=ACMYellow!40] [below = 0.15cm of M] (K) {Host OS kernel};	
	\node[main node, text width=0.7cm, align=center, minimum size=0.6cm,fill=ACMGreen!40] [left = 0.9cm of L2.center] (CF2) {\textbf{CP}\\\textbf{FILE}};
	\node[main node, text width=3.6cm, align=center, minimum size=0.5cm,fill=ACMOrange!40] [above = 0.25cm of SC2] (LC2) {C library};

	\node[main node, text width=1.5cm, align=center, minimum size=0.5cm,fill=ACMOrange!40] [above right = 0.25cm and 0.2cm of SC1.north] (LC12) {C library};

	\node[main node, text width=1.5cm, align=center, minimum size=0.5cm,fill=ACMOrange!40] [above left = 0.25cm and 0.2cm of SC1.north] (LC11) {C library};
	\node[main node, text width=3.6cm, align=center, minimum size=1.0cm,fill=ACMOrange!40] [above =-\pgflinewidth of LC2] (S3) {Microservice};

	\node[main node, text width=1.5cm, align=center, minimum height=1.0cm, fill=ACMOrange!40] [above =-\pgflinewidth of LC12] (S2)
        {\footnotesize{}Component~2};

	\node[main node, text width=1.5cm, align=center, minimum height=1.0cm,fill=ACMOrange!40] [above =-\pgflinewidth of LC11] (S1) {\footnotesize{}Component~1};

	\draw[dashed, thin] ([xshift=-0.1cm, yshift=0.1cm]S1.north west) rectangle ([xshift=0.1cm, yshift=-0.7cm]LC11.south east);

	\draw[dashed, thin] ([xshift=-0.1cm, yshift=0.1cm]S2.north west) rectangle ([xshift=0.1cm, yshift=-0.7cm]LC12.south east);
	\draw[dashed, thin] ([xshift=-0.1cm, yshift=0.1cm]S3.north west) rectangle ([xshift=0.1cm, yshift=-0.7cm]LC2.south east);

	\draw[dashed, thin] ([xshift=-0.2cm, yshift=0.2cm]S1.north west) rectangle ([xshift=0.8cm, yshift=0.1cm]H1.south east);
	
	\draw[dashed, thin] ([xshift=-0.2cm, yshift=0.2cm]S3.north west) rectangle ([xshift=0.8cm, yshift=0.1cm]H2.south east);

	\node[main node, text width=0.5cm, align=center, minimum size=0.3cm,fill=ACMGreen!40,font=\footnotesize] [below =0.05cm of S2.center] (SRC) {\emph{src}};

	\node[main node, text width=0.5cm, align=center, minimum size=0.3cm,fill=ACMGreen!40,font=\footnotesize] [above=2.0cm of CF2] (DST) {\emph{dst}};


	\draw [-stealth, double, thick] (SRC.east) -| ([xshift=1.3cm]SRC.east) |- (CF2.west);

	\draw [-stealth, double, thick] (CF2.north) -| (DST.south);

	\draw[very thick, densely dashdotted, ACMRed] ([xshift=-0.3cm, yshift=0.3cm]S1.north west) rectangle ([xshift=0.9cm, yshift=-0.1cm]H1.south east);
	
	\draw[very thick, densely dashdotted, ACMRed] ([xshift=-0.3cm, yshift=0.3cm]S3.north west) rectangle ([xshift=0.9cm, yshift=-0.1cm]H2.south east);

	\node[align=center, minimum height=0.5cm,very thick, densely dashdotted, ACMRed, draw=none] [above left =0.2cm and 1cm of S2] (CF) {\textbf{\coffer}};

	\node[align=center, minimum height=0.5cm,dashed, thin, draw=none] [above right =-0.5cm and 4.5cm of CF] (IS) {\smash{\textbf{isolation}}\vphantom{Coffer}};	

	\draw [-, thick, dashdotted, ACMRed,] ([xshift=0.0cm]CF.east) to  ([xshift=0.6cm]CF.east);

	\draw [-, thick, dashed] ([xshift=6.0cm]CF.east) to  ([xshift=6.6cm]CF.east);
	
		\node[align=center, minimum height=0.5cm] at ([xshift=0.085cm]$(H1.south)!0.5!(H2.south)$) (A) {\myc{A}};

		\node[align=center, minimum height=0.5cm] [above left=0.02 and 0.15cm of H1.south west] (B) {\myc{B}};
	
		\node[align=center, minimum height=0.5cm] [above right=0.02 and 0.15cm of H2.south east] (B) {\myc{B}};

		\node[align=center, minimum height=0.5cm] [above right=0.02 and 0.05cm of SC2.south east] (C) {\myc{C}};

		\node[align=center, minimum height=0.5cm] [above left=0.02 and 0.05cm of SC10.south west] (2) {\myc{C}};

 	\end{tikzpicture}}
  \caption{\coffer architecture}
  \label{fig:arch}
\end{figure}

\section{\coffer Design}
\label{sec:design}


\coffers are a new virtualisation and compartmentalization abstraction for application components. Such components can often be co-located and exchange data, and \coffers isolate them with support for low-overhead data exchange using CHERI capabilities. The design of \coffers has the following features:

\mypar{Separate namespaces} Unlike containers, \coffers do \change{not} rely on a shared OS kernel for namespace isolation. They use capabilities to add a new userspace-level isolation boundary, moving OS kernel functionality from a privileged to an unprivileged layer. \coffers only use the host OS for execution contexts, synchronisation, and I/O, thus resembling~VMs\@.

\mypar{Bypassed communication} \coffers are mutually untrusted, but communication bypasses the host OS kernel for performance. They use capabilities for on-demand access to memory regions used for communication, without compromising neighbouring memory.

\mypar{Low-overhead isolation} \coffers use capabilities for low-overhead isolation of both process and program modules. For example, \coffers can isolate shared libraries with minimal changes to the calling interface.

\mypar{Compatibility} \coffers use CHERI's hybrid-cap mode. Capabilities are thus hidden from application code, which only needs changes to use new communication APIs.

\begin{figure}[tb]
  \centering
  \resizebox{0.9\linewidth}{!}{%
	\begin{tikzpicture}[->,>=stealth',shorten >=0pt,auto,node distance=1.5cm, thick, main node/.style={rectangle,draw, font=\footnotesize, minimum size=5mm}]
	\node[main node, text width=6.6cm,align=center, minimum height=0.3cm,fill=ACMOrange!20] (CL0)  {
		C library (musl)
	};

	\node[main node, text width=5.8cm, align=center, minimum height=0.01cm,,fill=ACMOrange!20, yscale=0.2] [below = 0.05cm of CL0] (CLA) {};

	\node[main node, text width=6.6cm, align=center, minimum height=0.01cm,,draw=none,yscale=0.2] [below = 0.05cm of CL0] (CL) {};

	\node[main node, text width=6.8cm, align=center, minimum height=1.3cm,,fill=ACMPurple!20] [below = 0.65cm of CL] (LOS) {};

	\node[main node, text width=6.6cm, align=center, minimum height=0.01cm,fill=ACMPurple!20,,yscale=0.2] [below = 0.1cm of LOS.north] (SC1) {};

	\node[main node, text width=6.8cm, align=center, minimum height=0.01cm,fill=ACMPurple!20,yscale=0.2] [below = 0.05cm of LOS] (HC) {};

	\node[main node, text width=7.0cm, align=center, minimum height=1.3cm,fill=ACMBlue!20] [below = 0.65cm of HC] (M) {};

	\node[main node, text width=6.8cm, align=center, minimum height=0.01cm,fill=ACMLightBlue!20,,yscale=0.2] [below = 0.1cm of M.north] (HCM) {};


        \node[main node, text width=2.0cm, align=center, minimum height=0.4cm,,fill=ACMLightBlue!20] [above = -0.02cm of M.center] (CAPM) {\emph{CAP control}};

	\node[text width=1.5cm, align=center, minimum height=0.4cm,font=\footnotesize] [below = 0.05cm of CAPM] (CT) {\textbf{\monitor}};

	\node[main node, text width=2.0cm, align=center, minimum height=0.4cm,,fill=ACMLightBlue!20] [right = 0.15cm of CAPM] (TH) {\emph{threads}};

	\node[main node, text width=2.0cm, align=center, minimum height=0.4cm,,fill=ACMLightBlue!20] [left = 0.15cm of CAPM] (D) {\emph{disk I/O}};

	\node[main node, text width=2.0cm, align=center, minimum height=0.4cm,,fill=ACMLightBlue!20] [below = 0.05cm of TH] (TR) {\emph{time/r}};

	\node[main node, text width=2.0cm, align=center, minimum height=0.4cm,fill=ACMLightBlue!20] [below = 0.05cm of D] (N) {\emph{net I/O}};

	\node[main node, text width=2.0cm, align=center, minimum height=0.4cm,,fill=ACMPurple!10] [above = -0.02cm of LOS.center] (NET) {\smash{\emph{namespace}}\vphantom{/dev/cf}};

	\node[main node, text width=1.5cm, align=center, minimum height=0.4cm,,fill=ACMPurple!10] [right = 0.25cm of NET] (CF) {/dev/cf};

	\node[main node, text width=1.5cm, align=center, minimum height=0.4cm,,fill=ACMPurple!10] [left = 0.25cm of NET] (ST) {\smash{\emph{storage}}\vphantom{/dev/cf}};

	\node[main node, text width=1.5cm, align=center, minimum height=0.4cm,,fill=ACMPurple!10] [below = 0.05cm of ST] (VNS) {\smash{\emph{network}}\vphantom{/dev/cf}};

	\node[text width=1.5cm, align=center, minimum height=0.4cm,,font=\footnotesize] [below = 0.05cm of NET] (LT) {Library OS};

	\node[main node, text width=1.5cm, align=center, minimum height=0.4cm,,fill=ACMPurple!1, dashed] [below = 0.05cm of CF] (CFX) {\emph{Init}};

	\node[main node, text width=3.3cm, align=center, minimum height=0.01cm,,fill=ACMOrange!20] [above right = 0.05cm and 0cm of CL0.north west] (APP) {Program/library};

	\node[main node, text width=3.3cm, align=center, minimum height=0.01cm,,fill=ACMOrange!20] [above left = 0.05cm and 0cm of CL0.north east] (LIB) {Shared libraries (.so)};

	\draw[dashed, thin] ([xshift=-0.1cm, yshift=0.1cm]APP.north west) rectangle ([xshift=0.1cm, yshift=-0.1cm]CL.south east);

	\draw[dashed, thin] ([xshift=-0.2cm, yshift=0.2cm]APP.north west) rectangle ([xshift=0.1cm, yshift=-0.1cm]HC.south east);

	\node[font=\footnotesize, text width=3.3cm, align=center, minimum height=0.01cm,fill=ACMLightBlue!20] [above = 0.05cm of M.north] (HCT) {\textbf{hostcall via \texttt{CINVOKE}}};

	\node[font=\footnotesize, text width=3.3cm, align=center, minimum height=0.01cm,,fill=ACMPurple!20] [above = 0.05cm of LOS.north] (SCT) {\textbf{syscall via \texttt{CINVOKE}}};


	\draw [<->, thick] ([xshift=2.0cm]HCM.north) to  ([xshift=2.0cm]HC.south);

	\draw [<->, thick] ([xshift=2.75cm]HCM.north) to  ([xshift=2.75cm]HC.south);

	\draw [<->, thick] ([xshift=-2.0cm]HCM.north) to  ([xshift=-2.0cm]HC.south);

	\draw [<->, thick] ([xshift=-2.75cm]HCM.north) to  ([xshift=-2.75cm]HC.south);

	\draw [<->, thick] ([xshift=2.0cm]SC1.north) to  ([xshift=2.0cm]CL.south);

	\draw [<->, thick] ([xshift=2.75cm]SC1.north) to  ([xshift=2.75cm]CL.south);

	\draw [<->, thick] ([xshift=-2.0cm]SC1.north) to  ([xshift=-2.0cm]CL.south);

	\draw [<->, thick] ([xshift=-2.75cm]SC1.north) to  ([xshift=-2.75cm]CL.south);

	\draw[very thick, densely dashdotted, ACMRed] ([xshift=-0.3cm, yshift=0.3cm]APP.north west) rectangle ([xshift=0.2cm, yshift=-0.60cm]HC.south east);

		\node[align=center, minimum height=0.2cm] [right=1.2cm of SCT.center] (C1) {\myc{1}};

		\node[align=center, minimum height=0.2cm] [right=1.2cm of HCT.center] (C2) {\myc{2}};

		\node[align=center, minimum height=0.2cm,ACMRed, densely dashdotted] [above right=-0.3cm and 3.25cm of HCT.center] (A) {\myc{A}};

		\node[align=center, minimum height=0.2cm, dashed] [above right=-0.3cm and 3.15cm of SCT.center] (B) {\myc{B}};

		\node[align=center, minimum height=0.1cm, dashed] [left= -0.05cm of CLA.west] (C) {\myc{C}};
 	\end{tikzpicture}}
  \caption{Anatomy of a \coffer}\label{fig:arch2}
\end{figure}

\subsection{Architecture overview}
\label{sec:design:architectire}

\F\ref{fig:arch} shows the architecture of \coffers. Each \coffer~\myc{A} is an application component, such as a process or library, and has three parts: (i)~program binaries and their libraries; (ii)~a standard C library; and (iii)~a library OS\@.

\coffers add two new isolation boundaries, enforced through capabilities. The \emph{\monitor boundary} \myc{B} separates the \emph{\monitor} from all \coffers, and \coffers from each other. The \monitor is responsible for the lifecycle and isolation of \coffers, allows safe communication between them, and provides other primitives that cannot be implemented inside the unprivileged library OS (\eg storage and networking I/O, time, threading and synchronisation). It has access to the memory of all \coffers, but not the other way around.

The \emph{Program boundary} \myc{C} separates programs from the library OS that provides them the namespace for all OS primitives. A single library OS instance can thus host multiple, mutually-isolated programs with their own code and data (left-most \coffer in \F\ref{fig:arch}).

These isolation boundaries are enforced by CHERI capabilities; compartmentalized content cannot access memory beyond its boundary, except through the controlled interfaces described next. Finally, there is a classical separation from the host OS, using CPU rings and MMU-based isolation.

\begin{table*}[tb]
  \centering
  \footnotesize
    \caption{\coffer API}
    \renewcommand{\arraystretch}{0.8}
    \begin{tabular}{m{1.0cm}m{9.5cm}m{5.5cm}}
      \toprule
      \textbf{Type} & \textbf{API function}     & \textbf{Description}  \\
      \midrule
\multirow{1}{*}{\emph{Creation}} & \texttt{\textbf{cp\_cvm\_make}(cp\_config\_t *cfg, char *libos, char *disk.img, int argc, char *argv[])}
 		&  Create new \coffer\\

      \midrule

\multirow{5}{*}{\emph{\acs{file}}} & \texttt{\textbf{cp\_file\_make}(char *key, size\_t key\_size, void *addr, size\_t size)}
 		&  Make \acs{file} for buffer \texttt{addr} \& publish with \texttt{key}\\
& \texttt{\textbf{cp\_file\_destroy}(int file)}
 		&  Destroy \acs{file}\\
& \texttt{\textbf{cp\_file\_get}(char *key, size\_t key\_size)}
 		&  Get \acs{file} with \texttt{key} from another \coffer\\
& \texttt{\textbf{cp\_file\_read,cp\_file\_write}(int file, char *key, size\_t key\_size)}
 		&  Read/write data via \acs{file} \texttt{file}\\
& \texttt{\textbf{cp\_file\_wait,cp\_file\_notify}(int file)}
 		&  Wait/notify signal via \acs{file} \texttt{file}\\

\midrule

\multirow{4}{*}{\emph{\acs{call}}} & \texttt{\textbf{cp\_call\_make}(char *key, size\_t key\_size, void *func)}
 		&  Make \acs{call} for \texttt{func} \& publish with \texttt{key}\\
& \texttt{\textbf{cp\_call\_destroy}(int call)}
 		&  Destroy previously created \acs{call}\\
& \texttt{\textbf{cp\_call\_get}(char *key, size\_t key\_size)}
 		&  Get \acs{call} with \texttt{key} from another \coffer\\
& \texttt{\textbf{cp\_call}(int call, bool async, void *arg, size\_t size)}
 		&  Call \acs{file} \texttt{call} with arguments\\

\midrule

\multirow{6}{*}{\emph{\acs{stream}}}& \texttt{\textbf{cp\_stream\_make}(char *key, size\_t key\_size)}
 		&  Make \acs{stream} \& publish with given \texttt{key}\\
& \texttt{\textbf{cp\_stream\_destroy}(int stream)}
 		&  Destroy \acs{stream}\\
& \texttt{\textbf{cp\_stream\_get}(char *key, size\_t key\_size)}
 		&  Get \acs{stream} with \texttt{key} from another \coffer\\
& \texttt{\textbf{cp\_stream\_send}(int stream, void *buf, size\_t size)}
 		&  Send buffer through \acs{stream}\\
& \texttt{\textbf{cp\_stream\_recv}(int stream, long id, void *buf, size\_t size)}
 		&  Post buffer to receive through \acs{stream}.\\
& \texttt{\textbf{cp\_stream\_poll}(int stream, long *id, size\_t nid, int timeout)}
 		&  Poll for data on receive buffers of \acs{stream}\\

\bottomrule
    \end{tabular}
    \label{tab:api}
  \end{table*}


\subsection{Isolation boundaries}
\label{sec:design:coffer}



We now describe how \coffer{} are isolated in more detail~(see~\F\ref{fig:arch2}). Each program compartment contains the code and data of its binary, its dependencies (shared libraries), and the standard C library; the \coffer also contains the library OS, which provides the OS functionality.

 

Isolation boundaries are enforced by giving each its own default CHERI capabilities using the \texttt{pcc} and \texttt{dcc} registers~(see~\S\ref{sec:background:cheri}) with non-over\-lap\-ping address ranges; compartmentalized code thus cannot load, store or jump into memory outside that granted by the capabilities that it holds. To allow \myc{1} program $\rightarrow$ libOS and \myc{2} libOS $\rightarrow$ \monitor calls, \coffers use extra capabilities that grant controlled access to functions outside the respective compartment.

\coffers need to implement the equivalent of user/kernel separation using CHERI capabilities in userspace. When loading a program, a set of capabilities is therefore given to the syscall handler functions of the library OS. The standard C library uses these capabilities to invoke system calls on the library OS through the \texttt{CInvoke} instruction, while the rest of the application remains capability-unaware. The library OS has full access to the programs that it manages. 

\coffers also need to implement the equivalent of guest/host (or VM/hypervisor) separation using CHERI capabilities in userspace. When creating a \coffer, the \monitor installs capabilities to its own host system call handlers on the new library OS instance; in turn, the library OS uses \texttt{CInvoke} to invoke \monitor operations.

\subsection{Creation and communication API}
\label{sec:design:api}

\coffers combine compatibility and flexibility when isolating cloud services. They support the execution of complete application components using a process isolation abstraction, but also that of individual library components. 

\T\ref{tab:api} shows the \coffer API. New \coffers are created by \texttt{cp\_cvm\_make()}; similar to \texttt{fork()}/\texttt{exec()}, it accepts a disk image file, a program binary to load into the \coffer, and a function in that binary to launch.
If a \coffer isolates a standalone library, \texttt{cp\_call()} invokes functions in the library.


\label{sec:design:shared}

\coffers use CHERI capabilities for efficient inter-\coffer communication. The \monitor exchanges an initial set of capabilities between \coffers{} to allow communication.

\mypar{\acs{file}} This primitive introduces a file-like API to access memory from another \coffer at arbitrary granularity; the use of capabilities in \acs{file} permits bypassed access to memory without repeated mediation by the \monitor.

A \emph{donor} \coffer registers a memory region with the \monitor to share with other \coffers via \texttt{cp\_file\_make()}; a \emph{recipient} \coffer calls \texttt{cp\_file\_get()} with the same key to obtain access. The \coffers then access data in the memory region via \texttt{cp\_file\_read}/\texttt{write()}. Internally, the library OS uses capability-aware code to copy data directly between the \coffers (using \texttt{capcpy}; see \cref{sec:impl}).

To support asynchronous data transfers, \texttt{cp\_file\_wait()} and \texttt{cp\_file\_notify()} allow callers to wait for and notify events on a \acs{file}, respectively. Finally, the donor \coffer calls \texttt{cp\_file\_destroy()} to destroy it, revoking all access.


\mypar{\acs{call}} This primitive invokes functions outside the calling \coffer, \eg a callback function in the library OS, or a function in a shared library. \coffers manage \acsp{call} as follows: \texttt{cp\_call\_make()} registers a function in the donor that recipients can look up using \texttt{cp\_call\_get()} and then call with \texttt{cp\_call()}. The call is received by the \monitor, which creates a new thread in the donor's \coffer, sets it to execution to the target function with given arguments and, optionally, waits for its completion, based on the \texttt{async} argument.


\mypar{\acs{stream}} By composing the \acsp{file} and \acsp{call} APIs, it is possible to construct more complex communication mechanisms. For example, we have built a stream-oriented API for inter-\coffer communication in which the sender does not need to know where data is copied.

A recipient \coffer calls \texttt{cp\_stream\_recv()} to register buffers for incoming messages (internally, a list of \acsp{file}); a sender \coffer calls \texttt{cp\_stream\_send()} to copy data into any of the buffers available in the recipient. The recipient is then informed of data transfers when calling \texttt{cp\_stream\_poll()}.

\subsection{Capability management}

The use of CHERI capabilities introduces two problems that \coffers must avoid: avoiding the need for application code to become capability-aware and performance problems when revoking capabilities.

As explained in \cref{sec:background:cheri}, making an application fully capability-aware requires code changes. The design of \coffers avoids this by limiting the use of capability-aware code to a small portion of the standard C library, the library OS and the \monitor, which explicitly handle the \acsp{file} and \acsp{call} abstractions through syscall trampolines.

In the \coffer design, we want to avoid centralized trusted mechanisms for capability revocation~(see~\cref{sec:background:cheri}), as this goes against our goal of minimizing overheads and TCB size. Therefore, only the \monitor is permitted to store CHERI capabilities in memory: all capabilities that are passed by the \monitor to \coffers have the \texttt{CAP\_STORE} permission withheld. Instead of having to perform expensive garbage collection, revocation can now be done by clearing a small number of capability registers. This can be done efficiently when programs call the \coffer API to avoid interrupting execution.

\section{Implementation}
\label{sec:impl}


Next, we report implementation details of \coffers on the CHERI RISC-V64 platform. Our implementation consists of 5,200~lines of C code and 100~lines of assembly for the \monitor, and 1,800~lines of C code and 200~lines of assembly for the Init service, the Hostcall interface and CAP Devices. It uses the Linux Kernel Library~(LKL)~v4.17.0~\cite{lkl-src} as the library OS and the musl standard C library~v1.2.1~\cite{musl}. As the host OS kernel, we use CheriBSD~\cite{cheribsd}.


\subsection{\coffer lifecycle}
\label{sec:impl:lifecycle}
  
\mypar{Initialisation} The boot process of a \coffer is trigged by the \monitor. It receives a deployment configuration for the \coffer, which includes the heap size, the disk image location, the permitted interfaces, etc. It also defines the version and location of an Init service (see below) and the library OS binaries. The \monitor first allocates memory for the \coffer binary, stack and heap. It also allocates memory for the thread stack pool. Our implementation of \coffers cannot change the size of heap and stack at runtime, but this is a minor limitation given the size is in terms of virtual memory, and is only committed to physical memory on demand. Just as cloud providers prefer re-instantiating VMs over the use of memory ballooning, we expect large resource size changes to re-instantiate \coffers.

All threads must be created inside a compartment's memory, thus the \monitor pre-allocates memory for future thread stacks. After that, the \monitor deploys the image of the Init service into the \coffer and spawns the initial thread in the context of the \coffer. This thread prepares the hostcall callback tables, and enters the \coffer via the \texttt{CInvoke}-based interface created by the \monitor.

The Init service (see~\F\ref{fig:arch2}) is responsible for initializing all components at deployment, and creates the communication interface between the library OS and the host system. It is part of the library OS isolation layer, which means that it can access the memory of the application component. It initialises the library OS, builds the syscall interface for the program (or library), deploys its binary and calls the entry function (\eg \texttt{c\_start()}). For an executable binary, it launches the program; for a library, the entry function initializes a \acs{stream} and registers the public library functions with the \monitor.


\mypar{Execution} \coffers use the Linux kernel library~(LKL)~\cite{lkl-src} as a library OS that provides a Linux-compatible environment. LKL processes system calls and requests the host OS kernel to perform actions as needed.

LKL's storage and networking backends implement lean interfaces for hardware I/O devices: disk I/O has three hostcalls (\texttt{disk\_read}/\texttt{write()}, \texttt{disk\_getsize()}); networking uses only \texttt{net\_read}/\texttt{write()}. The \texttt{disk\_read}/\texttt{write} functions are applied to a file descriptor of the disk image; the network functions are invoked on a TAP device. The remaining functions in the hostcall interface are straightforward: they offer support for time and timer functions, debug output, threading and locking, and management of CAP Devices~(see~\cref{cap_impl}).

\mypar{Threading} For simplicity, \coffers use a 1-to-1 threading model. When a \coffer creates a thread, the \texttt{pthread} library requests an execution context from LKL, which in turn, requests a new thread from the host OS kernel. This requires the integration of the \texttt{pthread} implementations inside the \coffer and the host---both must maintain their own thread-local stores, pointers to \texttt{thread\_structs}, etc.

When LKL requests a thread, it prepares a structure with an address of the entrance function, and a pointer to the arguments. This is passed to the host OS kernel, and the \monitor creates a new thread with the provided arguments: it allocates a stack for the thread from the thread stack pool, pre-allocated at boot. After that, the new thread is ready to enter the \coffer using \texttt{CInvoke} and capabilities are created by the hostcall interface.  Prior to entering, the \monitor switches the thread pointer~\texttt{tp} register. Inside a \coffer, threads have LKL TP values; when processing hostcalls, they have host ones.

\begin{figure}[tb]
  \centering
  \resizebox{.9\linewidth}{!}{
	\begin{tikzpicture}[->,>=stealth',shorten >=1pt,auto,node distance=1.5cm, thick,main node/.style={rectangle,draw, font=\Huge,minimum size=10mm}]
	\node[dashed, main node, text width=9.0cm,align=left, minimum width=1.5cm] (P1) {
	$\mathit{SETUP}:$\\
	$\ \ \ \ \texttt{SC\ RET.seal}$\\
	$\ \ \ \ \texttt{SC\ MON.DDC.seal}$\\
	$\mathit{CALL}(Init, arg1, arg2):$\\
	$\ \ \ \ \texttt{CSeal\ ENTRY}$\\
	$\ \ \ \ \texttt{\$a0=arg1}$\\
	$\ \ \ \ \texttt{\$a1=arg2}$\\
	$\ \ \ \ \texttt{\$t0=ID\_INIT}$\\
	$\ \ \ \ \texttt{\$ddc=COMP.DDC}$\\
	$\ \ \ \ \texttt{CInvoke\ ENTRY.seal}$\\
	$\ $\\
	$\mathit{RET}:$\\
	$\ \ \ \ \texttt{\$ddc=MON.DDC}$\\
	};

	\node[dashed, main node, text width=15.5cm,align=left, minimum width=1.5cm] (P2) [right= 1.5cm of P1] {
	$\mathit{ENTRY}:$\\
	$\ \ \ \ \texttt{JR\ CALL\_TABLE[\$t0]}$\\
	$ $\\
	$\ \ \ \ \texttt{LC\ MON.DDC.seal}$\\
	$\ \ \ \ \texttt{LC\ RET.seal}$\\	
	$\ \ \ \ \texttt{CInvoke\ RET.seal, MON.DDC.seal}$\\
	$\ \ \ \ \ \ \ \ \ \ \ \  \mathit{OR}: \ \ \ \ \ $\\
	$\ \ \ \ \texttt{CInvoke\ OCALL.seal, MON.DDC.seal}$\\
	};

	\draw [->,very thick] ([yshift=-2.4cm,xshift=-0.65cm]P1.east) to [out=30,in=180] node [below,font=\LARGE] {} ([yshift=2.5cm]P2.west);
	
	\draw [<-,very thick] ([yshift=-4.6cm,xshift=-7.0cm]P1.east) to [out=0,in=210] node [below,font=\LARGE] {} ([yshift=-3.0cm,xshift=0.8cm]P2.west);
\node[font=\Huge, text width=8.0cm, align=center] (P1T) at ([yshift=-1.25cm]P1.south) {Outer\ Compartment \\ (e.g.\@ \monitor)};

\node[font=\Huge, text width=8.0cm, align=center] (P2T) at ([yshift=-1.25cm]P2.south) {Inner\ Compartment \\ (e.g.\@ Init)};

	\draw [-,thick, dashed] ([yshift=6.0cm,xshift=0.75cm]P1.east) to node {} ([yshift=-6cm,xshift=0.75cm]P1.east);
	
 	\end{tikzpicture}
 	
}
  \caption{ICALL and OCALL implementation}
  \label{fig:icall}
\end{figure}
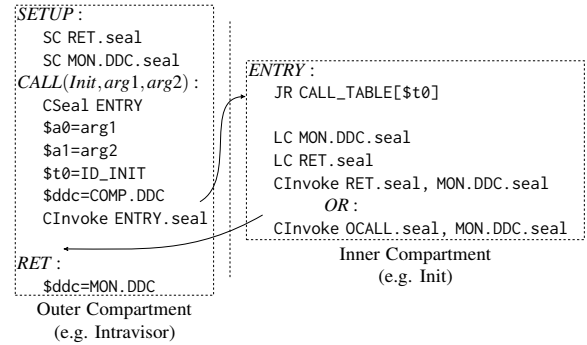

\subsection{Calls between nested compartments}
\label{sec:impl:nesting}


\coffers use the \texttt{CInvoke} instruction to call functions between isolation layers, both (i)~from an outer to an inner layer~(ICALL), \eg when the \monitor invokes Init; and (ii)~from an inner to an outer layer~(OCALL), \eg when performing a syscall or hostcall.

\texttt{CInvoke} takes two \emph{sealed} capabilities~(see~\cref{sec:background:cheri}) as arguments: (i)~one with a new Program Counter Capability~(\texttt{pcc}) value and another that points to a memory region that becomes accessible after the instruction execution. The \texttt{pcc} is replaced by the first unsealed capability; the second capability moves to the \texttt{ct6} (\texttt{C31}) register in the unsealed form.

Next, we explain how \texttt{CInvoke} is used to implement both ICALLs and OCALLs:

\mypar{ICALLs} \F\ref{fig:icall} shows the switching mechanism for ICALLs. In this example, the \monitor in the outer layer calls Init in the inner layer. To make the call, the caller prepares the first capability that points to the entry point inside the compartment. This capability, together with the corresponding data capability, defines the default capabilities of the inner compartment. Inside the compartment, these capabilities, \texttt{COMP.DDC} and \texttt{ENTRY.PCC} become \texttt{ddc} and \texttt{pcc}, respectively. While the \texttt{ENTRY.PCC} capability can be passed as the first argument of \texttt{CInvoke}, \texttt{COMP.DCC} must be loaded by the caller prior to switching (see~\F\ref{fig:icall}).


To return from the compartment or grant permission to invoke functions in the outer layer from the inner layer, further capabilities are needed: these are stored in memory by the \monitor before \texttt{CInvoke} is called, in a structure that we call the \emph{Affix}. They include a sealed \texttt{ddc} of the outer layer~(\texttt{MON.DDC.sealed}). Without this capability, the \monitor could not change \texttt{ddc} from the inner to the outer layer on return in order to access the \monitor's data. This capability can only be fetched from the inner layer---the accessible memory is restricted by the \texttt{ddc} of the inner layer.


The Affix also includes \texttt{RET.sealed} and \texttt{OCALL.sealed}, which are two sealed \texttt{pcc} capabilities to entry functions in the outer layer. The former is used to return from the compartment; the latter points to an entry function, which is used when the inner layer calls a function of the outer layer (\eg \texttt{print()}) and returns to continue execution inside the compartment. This is used for the syscall and hostcall interfaces. Capabilities in the Affix are created by the \monitor and stored on the stack and inside per-compartment private stores.

\myparr{OCALLs} share many similarities with ICALLS. The caller prepares a sealed capability of the return address. After the end of a function, the callee uses \texttt{CInvoke} and the execution of the caller continues from the desired address. Together with \texttt{CInvoke}, the callee passes the sealed capability \texttt{MON.DDC.seal}, which was passed originally inside the Affix. It is put into \texttt{ddc} after the function returns.


\subsection{Communication mechanisms}
\label{cap_impl}

The data sharing API between \coffers from \S\ref{sec:design:shared} is also based on capabilities. Data referenced by capabilities, however, can only be manipulated by capability-aware instructions, which do not exist in \change[hybrid-cap]{native} code. To resolve this issue, we \emph{mediate} the interaction between hybrid-cap code and capabilities using virtual devices called \emph{CAP Devices}.

The \acsp{file}, \acsp{call}, and \acsp{stream} primitives are implemented using character devices, which are created by the library OS and \monitor. A program can read/write from/to these devices, and the corresponding operations are performed by capability-aware code inside drivers.


This design has two advantages: (i)~despite its one memory copy, it is faster than traditional communication interfaces~(see~\S\ref{eval:microspeed}); and (ii)~it supports a simple mechanism to revoke capabilities. A remote \coffer can inform the \monitor of the revocation, which then requests the library OS to destroy the corresponding CAP Device. To revoke capabilities in pure-cap code, a \monitor would have to stop the \coffer execution and destroy capabilities manually.


\myparr{\acsp{file}} support regular POSIX file operations. In contrast to ordinary files, the content of \acsp{file} is not cached by the page cache, and read/write operations can be unaligned.

\begin{figure}[tb]
  \centering
  \begin{subfigure}[b]{1.0\linewidth}
  \centering
  \resizebox{.9\linewidth}{!}{
	\begin{tikzpicture}[->,>=stealth',shorten >=0pt,auto,node distance=1.0cm, thick,main node/.style={rectangle,draw, font=\normalsize,minimum size=5mm}]

	\node[main node, text width=1.0cm,align=center, minimum size=1.0cm] (DST)  {
	\emph{DST}
	};

	\node[main node, text width=1.0cm, align=center, minimum size=1.0cm] [right = 1cm of DST] (CAPDRV1) {\emph{CAP\\DRV}};

	\node[main node, text width=1.0cm, align=center, minimum size=0.5cm, dashed] [below = 0.8cm of CAPDRV1] (CAPCPY) {\emph{capcpy\vphantom{()}}};

	\node[main node, text width=1.0cm, align=center, minimum size=1.0cm] [right = 1cm of CAPDRV1] (CAPMNG) {\emph{CAP\\MNG}};

	\node[main node, text width=1.0cm, align=center, minimum size=1.0cm] [right = 1cm of CAPMNG] (CAPDRV2) {\emph{CAP\\DRV}};

	\node[main node, text width=1.0cm, align=center, minimum size=1.0cm] [right = 1cm of CAPDRV2] (SRC) {\emph{SRC}};

	\draw [-stealth, double, thick] (SRC.south) |- (CAPCPY.east);

	\draw [-stealth, double, thick] (CAPCPY.west) -| (DST.south);

	\draw [->,thick, dotted] ([yshift=0.3cm]DST.east) to [out=0,in=180] node [above,font=\footnotesize,yshift=0.2cm] {\emph{\_get("key")}} ([yshift=0.3cm]CAPDRV1.west);	

	\draw [->,thick] ([yshift=-0.3cm]DST.east) to [out=0,in=180] node [above,font=\footnotesize,yshift=-0.7cm] {\emph{\_read("key")}} ([yshift=-0.3cm]CAPDRV1.west);

	\draw [->,thick] (CAPDRV1.south) to [out=-90,in=90] node [above,font=\footnotesize,yshift=-0.7cm] {} (CAPCPY.north);

	\draw [->,thick, dotted] ([xshift=0.2cm]CAPDRV1.north) to [out=45,in=135] node [above,font=\footnotesize,yshift=0.0cm] {\emph{probe("key")}} ([xshift=-0.2cm]CAPMNG.north);	

	\draw [<-,thick, dotted] ([xshift=0.2cm]CAPDRV1.south) to [out=-45,in=-135] node [above,font=\footnotesize,yshift=0.0cm] {\emph{SRC\_CAP}} ([xshift=-0.2cm]CAPMNG.south);

	\draw [<-,thick, dotted] (CAPMNG.east) to [out=0,in=180] node [above,font=\footnotesize,yshift=0.5cm] {\emph{advertise("key", src, M)}} (CAPDRV2.west);	

	\draw [<-,thick, dotted] (CAPDRV2.east) to [out=0,in=180] node [above,font=\footnotesize,yshift=0.5cm, xshift=0.4cm] {\emph{\_make("key", *src)}} (SRC.west);

	\node[text width=5.0cm, align=center] (P1T) at ([yshift=-0.7cm,xshift=-1.2cm]CAPDRV2.south) {\texttt{ld.cap a0,\,SRC\_CAP}};

	\node[text width=2.0cm, align=center] (P1T) at ([yshift=-0.7cm,xshift=-5.9cm]CAPDRV2.south) {\texttt{sd a0,\,DST}};


	\draw [-, thin, dashed] ([yshift=0.5cm,xshift=0.0cm]$(CAPDRV2)!0.5!(SRC)$) to node {} ([yshift=-0.75cm,xshift=0.0cm]$(CAPDRV2)!0.5!(SRC)$);

	\draw [-, thin, dashed] ([yshift=1.75cm,xshift=0.0cm]$(CAPDRV2)!0.5!(SRC)$) to node {} ([yshift=1.25cm,xshift=0.0cm]$(CAPDRV2)!0.5!(SRC)$);

	\draw [-, thin, dashed] ([yshift=0.5cm,xshift=0.0cm]$(CAPDRV2)!0.5!(CAPMNG)$) to node {} ([yshift=-0.75cm,xshift=0.0cm]$(CAPDRV2)!0.5!(CAPMNG)$);

	\draw [-, thin, dashed] ([yshift=1.75cm,xshift=0.0cm]$(CAPDRV2)!0.5!(CAPMNG)$) to node {} ([yshift=1.25cm,xshift=0.0cm]$(CAPDRV2)!0.5!(CAPMNG)$);
	
	\draw [-, thin, dashed] ([yshift=0.5cm,xshift=0.0cm]$(CAPDRV1)!0.5!(CAPMNG)$) to node {} ([yshift=-0.75cm,xshift=0.0cm]$(CAPDRV1)!0.5!(CAPMNG)$);

	\draw [-, thin, dashed] ([yshift=1.75cm,xshift=0.0cm]$(CAPDRV1)!0.5!(CAPMNG)$) to node {} ([yshift=1.25cm,xshift=0.0cm]$(CAPDRV1)!0.5!(CAPMNG)$);

	\draw [-, thin, dashed] ([yshift=0.5cm,xshift=0.0cm]$(CAPDRV1)!0.5!(DST)$) to node {} ([yshift=-0.75cm,xshift=0.0cm]$(CAPDRV1)!0.5!(DST)$);

	\draw [-, thin, dashed] ([yshift=1.75cm,xshift=0.0cm]$(CAPDRV1)!0.5!(DST)$) to node {} ([yshift=1.25cm,xshift=0.0cm]$(CAPDRV1)!0.5!(DST)$);

	\node[text width=1.0cm, align=center] (P1T) at ([yshift=1.0cm,xshift=0cm]SRC.north) {\emph{Program}};

	\node[text width=1.0cm, align=center] (P1T) at ([yshift=1.0cm,xshift=0cm]DST.north) {\emph{Program}};

	\node[text width=2.0cm, align=center] (P1T) at ([yshift=1.0cm,xshift=0cm]CAPDRV2.north) {\emph{library OS}};

	\node[text width=2.0cm, align=center] (P1T) at ([yshift=1.0cm,xshift=0cm]CAPDRV1.north) {\emph{library OS}};

	\node[text width=2.0cm, align=center] (P1T) at ([yshift=1.0cm,xshift=0cm]CAPMNG.north) {\emph{\vphantom{y}\monitor}};
	
 	\end{tikzpicture}
	}
  \caption{\acsp{file}}
  \label{fig:capfile}
  \end{subfigure}
  \begin{subfigure}[b]{1.0\linewidth}
\centering
  \resizebox{.9\linewidth}{!}{
	\begin{tikzpicture}[->,>=stealth',shorten >=0pt,auto,node distance=1.0cm, thick,main node/.style={rectangle,draw, font=\normalsize,minimum size=5mm}]

	\node[main node, text width=1.0cm,align=center, minimum size=1.0cm] (DST)  {
	\emph{PY}
	};

	\node[main node, text width=1.0cm, align=center, minimum size=1.0cm] [right = 1cm of DST] (CAPDRV1) {\emph{CAP\\DRV}};

	\node[main node, text width=1.0cm, align=center, minimum size=1.0cm] [right = 1cm of CAPDRV1] (CAPMNG) {\emph{CAP\\MNG}};

	\node[main node, text width=1.0cm, align=center, minimum size=1.0cm] [right = 1cm of CAPMNG] (CAPDRV2) {\emph{CAP\\DRV}};

	\node[main node, text width=1.0cm, align=center, minimum size=1.0cm] [right = 1cm of CAPDRV2] (SRC) {\emph{reg.\vphantom{C}\\enc()}};

	\node[main node, text width=1.0cm, align=center, minimum size=0.5cm, dashed] [below = 0.4cm of SRC] (ENC) {\emph{enc()}};

	\draw [-stealth, double, thick] (CAPMNG.south) |- (ENC.west);

	\draw [->,thick, dashed] ([yshift=0.3cm]DST.east) to [out=0,in=180] node [above,font=\footnotesize,yshift=0.2cm] {\emph{\_get("key")}} ([yshift=0.3cm]CAPDRV1.west);	

	\draw [->,thick] ([yshift=-0.3cm]DST.east) to [out=0,in=180] node [above,font=\footnotesize,yshift=-0.7cm] {\emph{\_call("key", args)}} ([yshift=-0.3cm]CAPDRV1.west);	

	\draw [->,thick, dashed] ([yshift=0.3cm]CAPDRV1.east) to [out=0,in=180] node [above,font=\footnotesize,yshift=0.2cm] {\emph{probe("key")}} ([yshift=0.3cm]CAPMNG.west);	

	\draw [->,thick] ([yshift=-0.3cm]CAPDRV1.east) to [out=0,in=180] node [above,font=\footnotesize,yshift=-0.7cm] {\emph{call("key", args)}} ([yshift=-0.3cm]CAPMNG.west);

	\draw [<-,thick, dashed] (CAPMNG.east) to [out=0,in=180] node [above,font=\footnotesize,yshift=0.5cm] {\emph{advertise("key", \&enc, M)}} (CAPDRV2.west);	

	\draw [<-,thick, dashed] (CAPDRV2.east) to [out=0,in=180] node [above,font=\footnotesize,yshift=0.5cm, xshift=0.4cm] {\emph{\_make("key", \&enc)}} (SRC.west);

	\node[text width=3.0cm, align=center] (P1T) at ([yshift=-0.4cm,xshift=-0.25cm]CAPDRV2.south) {\texttt{CALL(enc, args)}};


	\draw [-, thin, dashed] ([yshift=0.5cm,xshift=0.0cm]$(CAPDRV2)!0.5!(SRC)$) to node {} ([yshift=-0.75cm,xshift=0.0cm]$(CAPDRV2)!0.5!(SRC)$);

	\draw [-, thin, dashed] ([yshift=1.75cm,xshift=0.0cm]$(CAPDRV2)!0.5!(SRC)$) to node {} ([yshift=1.25cm,xshift=0.0cm]$(CAPDRV2)!0.5!(SRC)$);

	\draw [-, thin, dashed] ([yshift=0.5cm,xshift=0.0cm]$(CAPDRV2)!0.5!(CAPMNG)$) to node {} ([yshift=-0.75cm,xshift=0.0cm]$(CAPDRV2)!0.5!(CAPMNG)$);

	\draw [-, thin, dashed] ([yshift=1.75cm,xshift=0.0cm]$(CAPDRV2)!0.5!(CAPMNG)$) to node {} ([yshift=1.25cm,xshift=0.0cm]$(CAPDRV2)!0.5!(CAPMNG)$);
	
	\draw [-, thin, dashed] ([yshift=0.5cm,xshift=0.0cm]$(CAPDRV1)!0.5!(CAPMNG)$) to node {} ([yshift=-0.75cm,xshift=0.0cm]$(CAPDRV1)!0.5!(CAPMNG)$);

	\draw [-, thin, dashed] ([yshift=1.75cm,xshift=0.0cm]$(CAPDRV1)!0.5!(CAPMNG)$) to node {} ([yshift=1.25cm,xshift=0.0cm]$(CAPDRV1)!0.5!(CAPMNG)$);

	\draw [-, thin, dashed] ([yshift=0.5cm,xshift=0.0cm]$(CAPDRV1)!0.5!(DST)$) to node {} ([yshift=-0.75cm,xshift=0.0cm]$(CAPDRV1)!0.5!(DST)$);

	\draw [-, thin, dashed] ([yshift=1.75cm,xshift=0.0cm]$(CAPDRV1)!0.5!(DST)$) to node {} ([yshift=1.25cm,xshift=0.0cm]$(CAPDRV1)!0.5!(DST)$);

	\node[text width=1.0cm, align=center] (P1T) at ([yshift=1.0cm,xshift=0cm]SRC.north) {\emph{Program}};

	\node[text width=1.0cm, align=center] (P1T) at ([yshift=1.0cm,xshift=0cm]DST.north) {\emph{Program}};

	\node[text width=2.0cm, align=center] (P1T) at ([yshift=1.0cm,xshift=0cm]CAPDRV2.north) {\emph{library OS}};

	\node[text width=2.0cm, align=center] (P1T) at ([yshift=1.0cm,xshift=0cm]CAPDRV1.north) {\emph{library OS}};

	\node[text width=2.0cm, align=center] (P1T) at ([yshift=1.0cm,xshift=0cm]CAPMNG.north) {\emph{\vphantom{y}\monitor}};

 	\end{tikzpicture}
	}
  \caption{\acsp{call}}
  \label{fig:capcall}
  \end{subfigure}
  \caption{Implementation of communication mechanisms}
\end{figure}
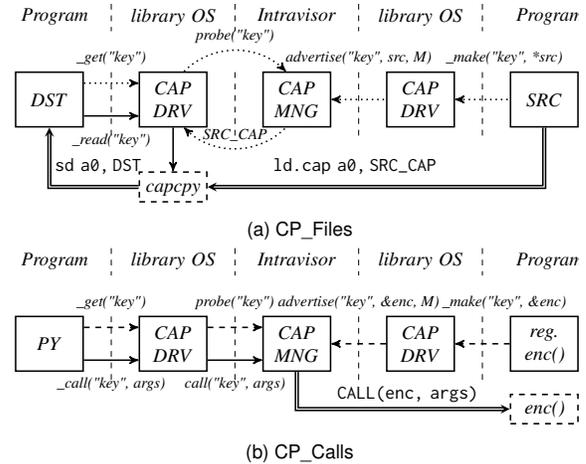

\F\ref{fig:capfile} shows the implementation. A donor \coffer advertises one or more memory regions defined by \emph{keys}, and a recipient \coffer probes the \monitor for a given key. The \monitor verifies the access control list and builds a CAP Device for the target \acs{file} (\eg \texttt{/dev/cf0}). For the donor \coffer to revoke access, it uses its own CAP Device to request revocation, and the \monitor, together with the library OS, destroy the \acsp{file} (\texttt{cf0}) driver along with its capabilities.
 
When the recipient \coffer issues a \texttt{cp\_file\_read()} call, the driver uses \emph{capcpy} to copy data. For \texttt{cp\_file\_read()}, it uses \texttt{ld.cap} to read data from a remote \coffer and store it via \texttt{sd}; a \texttt{cp\_file\_write()} does the reverse.

\mypar{\acsp{call}} To expose a function, a \coffer creates an ICALL entry and registers it with the \monitor~(see~\F\ref{fig:capcall}). The \monitor maintains a table of exported functions for each \coffer, called \coffer-RPCs. It consists of access control records with capabilities, name identifiers and permissions. Application components interact with the \coffer-RPCs via CAP Devices, a management interface~(\texttt{/dev/cf}), and the \monitor.

Any function can be invoked by \acsp{call} including ones inside the library OS. This enables the use of \acsp{call} as a notification mechanism between \acsp{file}. The donor blocks execution until the recipient \coffer reads data. It makes the \texttt{wait()} call with the driver, the driver puts the execution thread in the work queue and waits for the signal. Prior to blocking, it registers a wake-up \acs{call} with the \monitor. The recipient \coffer, in turn, finishes its operations with the \acsp{file}, and notifies the donor via this \acs{call}.

These basic operations can be composed to create higher-level protocols, and a single CAP Device can handle multiple memory regions. For example, for Redis~(see~\S\ref{sec:eval:redis}), we use a series of read/write operations with a single notification as well as batched reads with different capabilities.

\mypar{\acsp{stream}} In contrast to \acsp{file}, when sending data, the destination for \acsp{stream} is unknown, and \texttt{cp\_stream\_send()} only knows the source. Therefore, one side of the communication pre-registers one or more destination buffers via \texttt{cp\_stream\_recv()}, and uses \texttt{cp\_stream\_poll()} to block. The remote side uses \acs{call} to enter the remote compartment, atomically fetches one destination buffer from a pre-registered queue of buffers, and copies into this buffer data via \texttt{capcpy}. It then wakes up the poll queue and returns.

\mypar{Hostcall Interface} \change{The \monitor does not impose restrictions on the number of calls in the hostcall interface. For the LKL library OS, the \monitor provides 24~hostcalls for minimal operation. In addition, 2~hostcalls are necessary for disk I/O, 3 for network I/O, and 10 for the capability-based communication primitives.} 


\subsection{Capability revocation}
\label{sec:revoc}

Data transfers (\texttt{capcpy}) are performed by the drivers of CAP Devices without direct involvement of the \monitor, which enhances performance and reduces the TCB. This, however, means that the driver must have access to the capabilities provided by the donor. We do not consider the driver trusted, thus it may be compromised by an adversary who obtains access to capabilities and memory outside the \coffer after the end of a communication session. To mitigate against this threat, \coffers support a revocation mechanism. It guarantees that, once the donor \coffer revokes capabilities, they are destroyed, and a recipient \coffer cannot use them.

First, \coffers or communication capabilities are not created with the \texttt{PERMIT\_STORE\_CAP} permission. Code inside a \coffer thus cannot store capabilities to memory: it can load them, modify, create new capabilities, but it fails on \texttt{ST}. The communication capabilities are stored once by the \monitor, when the communication is established, and destroyed at the end. Second, the revoked capabilities in the CPU context are destroyed after a context switch by the host OS kernel.


\section{Security Analysis}
\label{sec:eval:sec_analysis}




According to our threat model from \S\ref{sec:background:threat_model}, an attacker can gain control over a \coffer. However, we guarantee that they cannot escape the compartment or access memory beyond its boundary due to the CHERI architectural properties~(see~\S\ref{sec:background:cheri}): the \texttt{ddc} and \texttt{pcc} capabilities always apply, are non-extensible, and are controlled by the \monitor.


Hybrid-cap code may be vulnerable to attacks that attempt to break execution flow. An adversary may inject capability-aware instructions (\eg \texttt{CLD}/\texttt{CSD}, \texttt{CInvoke}) to access data and code outside of the compartment. To do this, the adversary requires capabilities, which they cannot construct from the available data inside a \coffer.

To escape a compartment, an adversary must obtain appropriate capabilities. Each \coffer, however, only maintains a few capabilities: a compartment (i)~receives three sealed capabilities via Affixes, which can be inspected by an adversary but not unsealed to create new capabilities; and (ii)~may receive capabilities used by \acsp{file} and \acsp{stream}. These capabilities can be exploited by an adversary after gaining full control over the library OS. Since these are data capabilities, they cannot be used to create code capabilities, which are needed to escape the compartment. The adversary also cannot store these capabilities due to their permissions. Finally, they also cannot be exported outside of the compartment via the hostcall interface, because the interface does not handle capabilities and instead corrupts them.


Hybrid-cap code may contain security flaws, but an adversary cannot escape confinement, unless a flaw in the outer level provides them with unsealed capabilities. In our design, this is unlikely due to the \monitor's small TCB. The adversary cannot export or import capabilities via the hostcall interface or use them beyond a communication session. Vulnerable hybrid-cap code cannot abuse host system calls, escalate privileges or attack other \coffers, because the host OS kernel ignores all direct system calls from \coffers.

\change{\coffers are intra-process compartments that share micro-architectural state and rely on the correctness of the CHERI architecture, which does not have special mechanisms to prevent side-channel attacks. Nonetheless, there are plans for CHERI to include explicit compartment identifiers~(CIDs) in a future version of the architecture~\cite{watson2018capability}. This will ensure that sensitive micro-architectural state is appropriately tagged by each \coffer, similar to tagged TLB entries. This can be used to prevent attacks, such as training the branch predictor by one \coffer to direct speculative execution in another \coffer.}

\section{Evaluation}
\label{sec:eval}





We now explore the performance of \coffers and the proposed communication interfaces. We begin with an overview of our evaluation platforms and workloads (\cref{eval:ch}). We then compare the performance of applications deployed with \coffers and Docker containers (\cref{sec:eval:tier}). In \cref{sec:eval:redis}, we validate the efficiency of inter-\coffer communication mechanisms; in \cref{sec:eval:library}, we explore the use of \coffers for component compartmentalisation; \change[Finally,]{and in} \cref{eval:microspeed}, \change[concludes the section with]{we compare} inter-\coffer communication mechanisms with existing OS mechanisms.  \change{Finally, \cref{eval:teardown} explores the deployment performance of \coffers and Docker containers.}



\subsection{Experimental environment}
\label{eval:ch}



The CHERI architecture is under active development and, while ARM's Morello board with CHERI support has been announced~\cite{arm-morello}, it is unavailable at the time of writing.
Therefore, we use two \textbf{evaluation platforms}: (1)~a single-core FPGA-based CHERI implementation~\cite{project-we-use}; and (2)~a multi-core SiFive RISC-V implementation without CHERI support.

\mypari{FPGA CHERI} We synthesize an FPGA image from DARPA's CHERI FETT program~\cite{fett} (\texttt{agfi-026d853\-0\-0\-3\-d\-6c433a}), that ships with a single-core RISC-V64 CHERI system based on the FLUTE core (5-stage, in-order pipeline, running at 100\unit{MHz})~\cite{flute}, and execute it on AWS F1~\cite{aws_f1}. We use CheriBSD as the host OS kernel, compiled as a hybrid-cap system with LLVM v11.0.0 and cheribuild~\cite{cheribuild}.

The FPGA implementation enables a quantitative evaluation of \coffers, but has limitations:
(i)~it has a single-core CPU with low clock frequency;
(ii)~its peripheral devices, in particular storage devices, are emulated by the host;
and (iii)~DRAM latency is disproportionately low compared to the CPU clock speed.
As a consequence, we cannot realistically execute typical cloud workloads that are memory- and I/O-bound and use multiple CPU cores. We also cannot eliminate system noise by pinning tasks to separate cores.

\mypari{SiFive RISC-V} To avoid the abovementioned limitations, we also evaluate \coffers on a HiFive Unmatched RISC-V board~\cite{hifive}, which has 4~RISC-V64 (dual-issue, in-order) CPU cores running at 1.2\unit{GHz}. The CPU does not have CHERI support, and we instead \change[emulate it by using CLang-13 in ``sim'' mode, which replaces]{replace} all CHERI instructions with \change{their} native RISC-V \change[instructions]{versions}.
Our applications execute on Ubuntu~v20.04 with Linux~v5.11.0 and the RISC-V Docker port~\cite{risc-docker-src} with Alpine containers~\cite{alpine}.
Our IPC micro-benchmarks execute on FreeBSD~14, as the FPGA version uses CheriBSD, and we run them on both platforms.


This approach allows us to execute realistic cloud applications. \change{We run CHERI-equivalent code and data paths while remaining compatible with existing RISC-V platforms (\eg{} by replacing capability loads/stores with ordinary \texttt{ld}/\texttt{st} instructions, \texttt{CInvoke} with \texttt{jr}, etc.). Note that security is therefore not enforced.}

\label{sec:eval:use_cases}




\mypar{Application workloads} We explore \coffers using several cloud applications and micro-benchmarks to evaluate their performance and isolation requirements:

\mypari{NGINX/Redis (\cref{sec:eval:tier})} This is a two-tier microservice deployment that evaluates the YCSB benchmark~\cite{ycsb} using the NGINX~\cite{nginx} web server and the Redis~\cite{redis} key/value store. NGINX acts as an API gateway and translates REST requests into Redis queries. When co-located, these services have a substantial amount of communication between them.
We demonstrate that the \coffers{} interfaces, \acsp{file} and \acsp{stream}, significantly reduce overhead, using the SiFive platform to compare \coffers against a deployment using Docker containers~\cite{docker}.

\mypari{Redis (\cref{sec:eval:redis})} We execute a single-core Redis instance~\cite{redis} and measure the latency of fixed-size GET and SET operations, comparing sockets and the equivalent \coffer interface with \acsp{stream}.
This experiment validates our previous results by also comparing the FPGA and SiFive environments.

\mypari{Python/Library (\cref{sec:eval:library})} We measure the cost of using \coffers to isolate the components of a simple cryptographic application in Python, by deploying the Python runtime~\cite{cpython} and the PyCrypto cryptographic library~\cite{pycrypto} in mutually isolated \coffers that use the \acs{call} and \acs{file} interfaces to communicate.
This experiment runs on the FPGA environment.


\subsection{Multi-tier deployment with NGINX/Redis}
\label{sec:eval:tier}

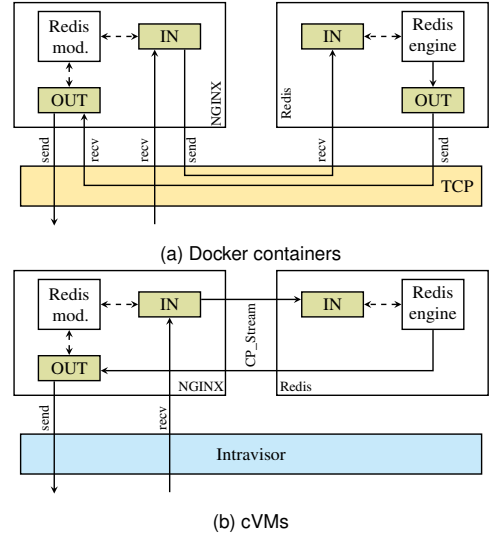
\begin{figure}[tb]
  \centering
  \begin{subfigure}[b]{.8\columnwidth}
    \centering
    \resizebox{0.95\linewidth}{!}{
	\begin{tikzpicture}[->,>=stealth',shorten >=0pt,auto,node distance=1.0cm, thick,main node/.style={rectangle,draw, font=\normalsize,minimum size=5mm}]

	\node[main node, text width=1.0cm,align=center, minimum size=0.5cm,fill=ACMGreen!40] (N_IN)  {IN};

	\node[main node, text width=1.0cm, align=center, minimum size=0.5cm,fill=ACMGreen!40] [right = 2cm of N_IN] (R_OUT) {IN};

	\node[main node, text width=1.0cm, align=center, minimum size=0.5cm,fill=ACMGreen!40] [below right = 0.75cm and 0.75cm of R_OUT] (R_IN) {OUT};

	\node[main node, text width=1.0cm, align=center, minimum size=1.0cm] [ right = 0.75cm of R_OUT] (ENG) {Redis engine};

	\node[main node, text width=1.0cm, align=center, minimum size=1.0cm] [ left = 0.75cm of N_IN] (MOD) {Redis mod.};

	\node[main node, text width=1.0cm, align=center, minimum size=0.5cm,fill=ACMGreen!40] [below left = 0.75cm and 0.75cm of N_IN] (N_OUT) {OUT};

	\node[main node, text width=9.0cm, align=center, minimum size=0.8cm, fill=ACMYellow!40] (TCP) at([yshift=-3cm]$(R_OUT)!0.5!(N_IN)$) {};

	\node[font=\normalsize] (TCP_T) at([xshift=-0.5cm]TCP.east) {TCP};

%
	\draw [-stealth, thick] ([xshift=0.3cm]N_IN.south) |- ([yshift=-0.2cm]TCP.north) -| ([xshift=-0.0cm]R_OUT.south);

	\draw [stealth-](R_IN.north) to [out=90, in=-90] (ENG.south);

	\draw [stealth-stealth, dashed](ENG.west) to  (R_OUT.east);

	\draw [stealth-stealth, dashed](N_OUT.north) to [out=90, in=-90] (MOD.south);

	\draw [stealth-stealth, dashed](MOD.east) to  (N_IN.west);

	\draw [-stealth, thick] ([xshift=0.0cm]R_IN.south) |- ([yshift=-0.4cm]TCP.north) -| ([xshift=0.3cm]N_OUT.south);

	\node[main node, text width=4.0cm, align=center, minimum size=2.5cm] [above left = 0.75cm and 0.5cm of TCP.north] (D1) {};

	\node[main node, text width=4.0cm, align=center, minimum size=2.5cm] [above right = 0.75cm and 0.5cm of TCP.north] (D2) {};


	\draw [-stealth, thick] ([xshift=-0.3cm,yshift=-3.5cm]N_IN.south) -- ([xshift=-0.3cm]N_IN.south);

	\draw [stealth-, thick] ([xshift=-0.3cm,yshift=-2.25cm]N_OUT.south) to [out=90, in=-90] node [above,font=\footnotesize,xshift=-0.2cm] (S1) {\rotatebox{90}{send}} ([xshift=-0.3cm]N_OUT.south);

	\node[font=\footnotesize] [right = 0.6cm of S1] (S2) {\rotatebox{90}{recv}};

	\node[font=\footnotesize] [right = 0.6cm of S2] (S3) {\rotatebox{90}{recv}};

	\node[font=\footnotesize] [right = 0.5cm of S3] (S4) {\rotatebox{90}{send}};

	\node[font=\footnotesize] [right = 2.2cm of S4] (S5) {\rotatebox{90}{recv}};

	\node[font=\footnotesize] [right = 2.0cm of S5] (S6) {\rotatebox{90}{send}};


	\node[font=\footnotesize] (DT1) at([xshift=-0.2cm,yshift=0.5cm]D1.south east) {\rotatebox{90}{NGINX}};

	\node[font=\footnotesize] (DT1) at([xshift=0.2cm,yshift=0.5cm]D2.south west) {\rotatebox{90}{Redis}};

 	\end{tikzpicture}
}
	
    \caption{Docker containers}\label{fig:t2:docker}
  \end{subfigure}
  \begin{subfigure}[b]{.8\columnwidth}
    \centering
    \resizebox{0.95\linewidth}{!}{
	\begin{tikzpicture}[->,>=stealth',shorten >=0pt,auto,node distance=1.0cm, thick,main node/.style={rectangle,draw, font=\normalsize,minimum size=5mm}]

	\node[main node, text width=1.0cm,align=center, minimum size=0.5cm,fill=ACMGreen!40] (N_IN)  {IN};

	\node[main node, text width=1.0cm, align=center, minimum size=0.5cm,fill=ACMGreen!40] [right = 2cm of N_IN] (R_OUT) {IN};

	\node[main node, draw=none, text width=1.0cm, align=center, minimum size=0.5cm] [below right = 0.75cm and 0.75cm of R_OUT] (R_IN) {};

	\node[main node, text width=1.0cm, align=center, minimum size=1.0cm] [ right = 0.75cm of R_OUT] (ENG) {Redis engine};

	\node[main node, text width=1.0cm, align=center, minimum size=1.0cm] [ left = 0.75cm of N_IN] (MOD) {Redis mod.};

	\node[main node, text width=1.0cm, align=center, minimum size=0.5cm,fill=ACMGreen!40] [below left = 0.75cm and 0.75cm of N_IN] (N_OUT) {OUT};

	\node[main node, text width=9.0cm, align=center, minimum size=0.8cm,fill=ACMLightBlue!40] (TCP) at([yshift=-3cm]$(R_OUT)!0.5!(N_IN)$) {\monitor};
%
	\draw [-stealth, thick] ([yshift=0.1cm]N_IN.east) |- ([yshift=0.1cm]R_OUT.west);


	\draw [stealth-stealth, dashed](ENG.west) to  (R_OUT.east);

	\draw [stealth-stealth, dashed](N_OUT.north) to [out=90, in=-90] (MOD.south);

	\draw [stealth-stealth, dashed](MOD.east) to  (N_IN.west);

	\draw [-stealth, thick] ([yshift=0.00cm]ENG.south) |- ([yshift=-0.05cm]N_OUT.east);

	\node[main node, text width=4.0cm, align=center, minimum size=2.5cm] [above left = 0.75cm and 0.5cm of TCP.north] (D1) {};

	\node[main node, text width=4.0cm, align=center, minimum size=2.5cm] [above right = 0.75cm and 0.5cm of TCP.north] (D2) {};


	\draw [-stealth, thick] ([xshift=-0.0cm,yshift=-3.5cm]N_IN.south) -- ([xshift=-0.0cm]N_IN.south);

	\draw [stealth-, thick] ([xshift=-0.3cm,yshift=-2.25cm]N_OUT.south) to [out=90, in=-90] node [above,font=\footnotesize,xshift=-0.2cm] (S1) {\rotatebox{90}{send}} ([xshift=-0.3cm]N_OUT.south);

	\node[font=\footnotesize] [right = 0.6cm of S1] (S2) {\rotatebox{90}{}};

	\node[font=\footnotesize] [right = 1.1cm of S2] (S3) {\rotatebox{90}{recv}};

	\node[font=\footnotesize] [above = 1.2cm of TCP] (CF) {\rotatebox{90}{\acs{stream}}};

	\node[font=\footnotesize] (DT1) at([xshift=-0.5cm,yshift=0.2cm]D1.south east) {\rotatebox{0}{NGINX}};

	\node[font=\footnotesize] (DT1) at([xshift=0.4cm,yshift=0.2cm]D2.south west) {\rotatebox{0}{Redis}};

 	\end{tikzpicture}
}
	
    \caption{\coffers}\label{fig:t2:coffer}
  \end{subfigure}
  \caption{Control/data flow in multi-tier deployment (NGINX/Redis)
  }
  \label{fig:t2}
\end{figure}

\begin{figure}[tb]
  \centering
  \begin{subfigure}[b]{.99\columnwidth}
    \centering
\centering
\begin{tikzpicture}
	\begin{axis}[
		legend columns=1,
		legend style={
    			draw=none,
    			fill=none,
		},
		legend cell align={left},
		legend style={at={(0.18,.40)},anchor=south, font=\footnotesize},
		width  = 1.0\linewidth,
		height = 5.0cm,
       xmin=0, xmax=3000,
       ymin=1, ymax=10,
     ]
\addlegendimage{empty legend}
\addlegendentry{\hspace{-0.4cm}\textbf{Containers}}
\addplot[
		each nth point={2},
		dash dot, ACMRed,  thick, mark=o,mark options={solid, scale=0.75},
		] table[x =x, y =HGETALL]{ycsb-dockers-50-HGETALL.txt};
\addlegendentry{HGETALL}

\addplot[
		each nth point={1},
		dash dot, ACMRed,  thick, mark=square,mark options={solid, scale=0.75},
		] table[x =x, y = ZRANGEBYSCORE]{ycsb-dockers-50-ZRANGEBYSCORE.txt};
\addlegendentry{ZRANGE}
 
       \addplot[
		each nth point={1},
		 dash dot, ACMRed,  thick, mark=star,mark options={solid, scale=0.75},
		] table[x =x, y = ZADD]{ycsb-dockers-50-ZADD.txt};
\addlegendentry{ZADD}

\addplot[
		each nth point={1},
		dash dot, ACMRed,  thick, mark=triangle,mark options={solid, scale=0.75},
		] table[x =x, y = HMSET]{ycsb-dockers-50-HMSET.txt};
\addlegendentry{HMSET}
 	\end{axis}
 	
	\begin{axis}[
		legend columns=1,
		legend style={
    			draw=none,
    			fill=none,
		},
		legend cell align={left},
		legend style={at={(0.86,.05)},anchor=south, font=\footnotesize},
		width  = 1.0\linewidth,
		height = 5.0cm,
       xmin=0, xmax=3000,
       ymin=1, ymax=10,
       xlabel near ticks,
       ylabel={50 percentile latency, ms},
       ylabel near ticks,
     ]
       \addplot[
		each nth point={1},
		ACMGreen,  thick, mark=o,mark options={solid, scale=0.75},
		] table[x =x, y =HGETALL]{ycsb-coffers-50-HGETALL.txt};

       \addplot[
		each nth point={1},
		ACMGreen,  thick, mark=square,mark options={solid, scale=0.75},
		] table[x =x, y = ZRANGEBYSCORE]{ycsb-coffers-50-ZRANGEBYSCORE.txt};

       \addplot[
		each nth point={1},
		 ACMGreen,  thick, mark=star,mark options={solid, scale=0.75},
		] table[x =x, y = ZADD]{ycsb-coffers-50-ZADD.txt};
 
       \addplot[
		each nth point={1},
		 ACMGreen,  thick, mark=triangle,mark options={solid, scale=0.75},
		] table[x =x, y = HMSET]{ycsb-coffers-50-HMSET.txt};

 	\end{axis}
\end{tikzpicture}
  \end{subfigure}
  \begin{subfigure}[b]{.99\columnwidth}
    \centering
\centering
\begin{tikzpicture}
	\begin{axis}[
		legend columns=1,
		legend style={
    			draw=none,
    			fill=none,
		},
		legend cell align={left},
		legend style={at={(0.18,.55)},anchor=south, font=\scriptsize},
		width  = 1.0\linewidth,
		height = 5.0cm,
       xmin=0, xmax=3000,
       ymin=1, ymax=10,
     ]
\addplot[
		each nth point={1},
		dash dot, ACMRed,  thick, mark=o,mark options={solid, scale=0.75},
		] table[x =x, y =HGETALL]{ycsb-dockers-95-HGETALL.txt};

\addplot[
		each nth point={1},
		dash dot, ACMRed,  thick, mark=square,mark options={solid, scale=0.75},
		] table[x =x, y = ZRANGEBYSCORE]{ycsb-dockers-95-ZRANGEBYSCORE.txt};

\addplot[
		each nth point={1},
		dash dot, ACMRed,  thick, mark=star,mark options={solid, scale=0.75},
		] table[x =x, y = ZADD]{ycsb-dockers-95-ZADD.txt};

\addplot[
		each nth point={1},
		dash dot, ACMRed,  thick, mark=triangle,mark options={solid, scale=0.75},
		] table[x =x, y = HMSET]{ycsb-dockers-95-HMSET.txt};
 	\end{axis}
	\begin{axis}[
		legend columns=1,
		legend style={
    			draw=none,
    			fill=none,
		},
		legend cell align={left},
		legend style={at={(0.18,.40)},anchor=south, font=\scriptsize},
		width  = 1.0\linewidth,
		height = 5.0cm,
       xmin=0, xmax=3000,
       ymin=1, ymax=10,
       xlabel near ticks,
       xlabel={Requests per second},
       ylabel={95 percentile  latency, ms},
       ylabel near ticks,
     ]
\addlegendimage{empty legend}
\addlegendentry{\hspace{-0.3cm}\textbf{\coffers}}
       \addplot[
		each nth point={1},
		ACMGreen,  thick, mark=o,mark options={solid, scale=0.75},
		] table[x =x, y =HGETALL]{ycsb-coffers-95-HGETALL.txt};
\addlegendentry{HGETALL}

       \addplot[
		each nth point={1},
		ACMGreen,  thick, mark=square,mark options={solid, scale=0.75},
		] table[x =x, y = ZRANGEBYSCORE]{ycsb-coffers-95-ZRANGEBYSCORE.txt};
\addlegendentry{ZRANGE}

       \addplot[
		each nth point={1},
		 ACMGreen,  thick, mark=star,mark options={solid, scale=0.75},
		] table[x =x, y = ZADD]{ycsb-coffers-95-ZADD.txt};
\addlegendentry{ZADD}
 
       \addplot[
		each nth point={1},
		 ACMGreen,  thick, mark=triangle,mark options={solid, scale=0.75},
		] table[x =x, y = HMSET]{ycsb-coffers-95-HMSET.txt};
\addlegendentry{HMSET}

 	\end{axis}
\end{tikzpicture}
  \end{subfigure}
  \caption{Multi-tier deployment performance (NGINX/Redis)} 
  \label{fig:eval:ycsb}
\end{figure}
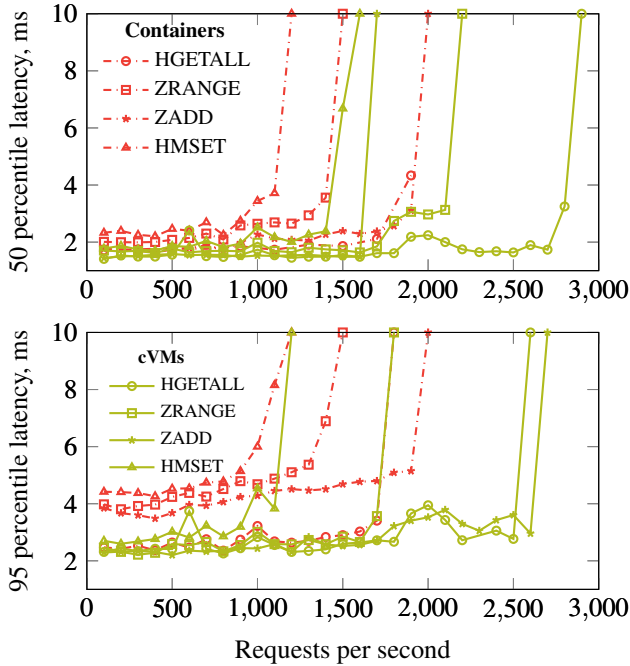


First, we compare the benefits of using \coffers when co-locating communicating components, compared to a traditional deployment with Docker containers~\cite{docker}.

The computational limitations of our FPGA and SiFive platforms make it unfeasible to execute a complete microservice benchmark suite such as DeathStarBench~\cite{deathstarbench}. Instead, we deploy a representative YCSB benchmark~\cite{ycsb} (workloadb; 1\unit{KB} records; read/update ratio of 95\%/5\%) on the SiFive platform with two-tiers: the NGNIX web server~\cite{nginx} acts as an API gateway that redirects incoming HTTP requests to the Redis key/value store~\cite{redis}, which acts as a cache for frequently used data. We use \emph{wrk2}~\cite{wrk2} to generate NGINX requests over a 1~GbE network, measuring the latency of different configurations (10~connections; 4~I/O threads).

%

The application components benefit from co-location due to the frequent interaction between the (NGINX) API gateway and its (Redis) cache. \cref{fig:t2} compares the Docker and \coffer deployments. Docker incurs multiple data copies between the components and the TCP/IP network stacks. As \cref{fig:t2:docker} shows, Redis copies values into a send buffer that is passed to the TCP/IP stack, which NGINX copies into an output buffer that is, in turn, passed to the client's network stack (for a total of 4~copies, including the kernel's TCP/IP stack).

\change{In contrast, \coffers reduce the number of copies. \cref{fig:t2:coffer} shows that the \acs{stream} primitive requires only 2~copies: Redis values are always copied directly into NGINX's output buffer. To support this optimization, NGINX and Redis must replace their use of sockets with \acsp{stream}. NGINX registers the output buffer with a \acs{stream}, and the \acs{stream} write in Redis uses capabilities to copy data directly into the output buffer, which NGINX can then send to the client.}


\Cref{fig:eval:ycsb} shows the median and 95\textsuperscript{th} percentile latencies for the 4~YCSB queries under various throughput regimes, comparing the baseline Docker deployment with \coffers.
We can see that \coffers are more efficient: they have lower latencies in all cases (20--40\% for median latency), and substantially higher throughput, with send latencies below 5\,ms (33--50\% for median latency).




\mypar{Conclusion} \change{In a typical deployment with multiple application components, \coffers can achieve isolation while lowering latencies and increasing throughput compared to containers. This performance gain is due to a reduced number of memory copies (via \acs{stream}), using fast calls to the capability-hiding TCB in \coffers (via \acs{call} within \acsp{stream}). Furthermore, \coffers{} come with a smaller TCB compared to containers. We also expect \coffers{} to outperform VMs because of VMs' higher overheads caused by memory virtualization (especially for memory-bound applications) and  communication mechanisms (\eg extra data copies by the guest OS and/or hypervisor, or cross-VM copies via PCIe with directly assigned devices).}

\subsection{Platform validation with Redis}
\label{sec:eval:redis}


We now validate our results by comparing the FPGA and SiFive platforms. We use Redis with a single connection that measures the latency of 1000~GET or SET operations with fixed-size keys (1\unit{byte}) and values (100\unit{bytes}). We use a simple client application that is co-located with the Redis instance. The baseline system uses separate processes and TCP/IP sockets; we use separate \coffers for each application and \acs{stream} for communication (similarly to \cref{sec:eval:tier}).






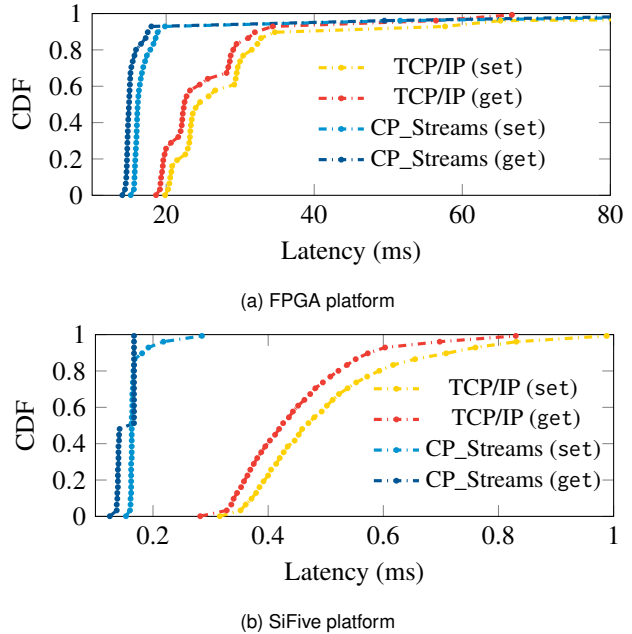
\begin{figure}[tb]
  \centering
  \begin{subfigure}[t]{1.0\linewidth}
    \centering
\begin{tikzpicture}
	\begin{axis}[
		legend columns=1,
		legend style={
    			draw=none,
		},
		legend style={at={(0.66,0.08)},anchor=south,draw=none},
		width  = 1.0\linewidth,
		height = 4cm,
      xmin=10000,xmax=80000,
       ymin=0, ymax=1,
       style={align=center}, ylabel={CDF},
       xlabel near ticks,
       xlabel={Latency (ms)},
       ylabel near ticks, 
			     xtick={
20000,
40000,
60000,
80000
},
    			  xticklabels={
20,
40,
60,
80
},xtick scale label code/.code={},
     ]
       \addplot[
		each nth point={32},
		dash dot, ACMYellow,  very thick, mark=*,mark options={solid, scale=0.25},
		] table[x =TCPSET, y =x]{redis-cdf.txt};
 \addlegendentry{\small{}TCP/IP (\texttt{set})}
 
        \addplot[
		each nth point={32},
		dash dot, ACMRed,  very thick, mark=*,mark options={solid, scale=0.25},
		] table[x =TCPGET, y =x]{redis-cdf.txt};
 \addlegendentry{\small{}TCP/IP (\texttt{get})}
 
         \addplot[
		each nth point={32},
		dash dot, ACMBlue,  very thick, mark=*,mark options={solid, scale=0.25},
		] table[x =CAPSET, y =x]{redis-cdf.txt};
 \addlegendentry{\small{}\acsp{stream} (\texttt{set})}

         \addplot[
		each nth point={32},
		dash dot, ACMDarkBlue,  very thick, mark=*,mark options={solid, scale=0.25},
		] table[x =CAPGET, y =x]{redis-cdf.txt};
 \addlegendentry{\small{}\acsp{stream} (\texttt{get})}
 
 \end{axis}
\end{tikzpicture}
   \caption{FPGA platform}\label{fig:eva:cdf1}
 \end{subfigure}
 
 \begin{subfigure}[t]{1.0\linewidth}
   \centering
   \begin{tikzpicture}
	\begin{axis}[
		legend columns=1,
		legend style={
    			draw=none,
    			fill=none,
		},
		legend style={at={(0.76,0.08)},anchor=south,draw=none},
		width  = 1.0\linewidth,
		height = 4cm,
       xmin=100,xmax=1000,
       ymin=0, ymax=1,
       style={align=center}, ylabel={CDF},
       xlabel near ticks,
       xlabel={Latency (ms)},
       ylabel near ticks, 
			     xtick={
200,
400,
600,
800,
1000
},
    			  xticklabels={
0.2,
0.4,
0.6,
0.8,
1
},xtick scale label code/.code={},
     ]
       \addplot[
		each nth point={32},
		dash dot, ACMYellow,  very thick, mark=*,mark options={solid, scale=0.25},
		] table[x =TCPSET, y =x]{sifive-redis-cdf.txt};
 \addlegendentry{\small{}TCP/IP (\texttt{set})}
 
        \addplot[
		each nth point={32},
		dash dot, ACMRed,  very thick, mark=*,mark options={solid, scale=0.25},
		] table[x =TCPGET, y =x]{sifive-redis-cdf.txt};
 \addlegendentry{\small{} TCP/IP (\texttt{get})}
 
         \addplot[
		each nth point={32},
		dash dot, ACMBlue,  very thick, mark=*,mark options={solid, scale=0.25},
		] table[x =CAPSET, y =x]{sifive-redis-cdf.txt};
 \addlegendentry{\small{}\acsp{stream} (\texttt{set})}

         \addplot[
		each nth point={32},
		dash dot, ACMDarkBlue,  very thick, mark=*,mark options={solid, scale=0.25},
		] table[x =CAPGET, y =x]{sifive-redis-cdf.txt};
 \addlegendentry{\small{}\acsp{stream} (\texttt{get})}
 \end{axis}
\end{tikzpicture}
\caption{SiFive platform}\label{fig:eva:cdf2}
  \end{subfigure}
  \caption{Latency CDF for Redis (platform validation)
  }\label{fig:eva:cdf}
\end{figure}



\Cref{fig:eva:cdf} shows the latency distribution of the GET and SET requests for all configurations.  The results indeed validate our observations from the multi-tier YCSB benchmark in \cref{sec:eval:tier}. \coffers{} exhibit lower latencies with less deviation on both platforms, compared to a native system with TCP/IP sockets: 90\% of \coffer requests take 14--19\unit{ms}; the baseline takes 19--35\unit{ms} on the FPGA platform. The SiFive platform supports the same conclusions, albeit with different absolute numbers. This is because the FPGA device runs at a lower clock frequency, and two processes must be co-scheduled on the same core (with both the baseline and \coffers{}).

\mypar{Conclusion} The \acs{stream} primitive in \coffers{} shows better performance on both the FPGA and SiFive platforms, achieving lower communication latencies across the whole throughput spectrum.  We thus conclude that our end-to-end evaluation in \cref{sec:eval:tier} is representative of how \coffers{} would perform on a real-world CHERI-enabled CPU\@. In \cref{eval:microspeed}, we re-validate this by comparing \coffers{} against IPC primitives on all platforms.

\subsection{Process compartmentalization with Python library}
\label{sec:eval:library}

Next, we explore the overhead of compartmentalizing a shared library with cryptographic operations in Python. In this case, we harden the security of a cloud application by mutually isolating the Python runtime and a native cryptographic module, PyCryptodome~\cite{pycrypto}. By using separate \coffers, we can safeguard the application against malicious interference by package managers~\cite{npm_trojan}, or protect the library against unauthorized access to its cryptographic keys~\cite{CVE-2014-0160}. 


Python creates \acsp{file} for the input/output buffers that it passes to the PyCryptodome library, and it uses \acs{call} to transfer control to the library, using the \acsp{file} as arguments. (The original version instead passes raw buffer pointers.) PyCryptodome then uses these \acsp{file} to read its input and encrypt/decrypt it into the output buffer(using AES-128. Finally, it uses \acs{call} to return execution to Python.


\begin{figure}[tb]
\centering
\begin{tikzpicture}
	\begin{axis}[
		legend columns=1,
		legend style={
    			draw=none,
    			fill=none,
		},
		legend cell align={left},
		legend style={at={(0.80,.05)},anchor=south, font=\scriptsize},
		width  = 1.0\linewidth,
		height = 4.5cm,
       xmin=16, xmax=8388608,
       ymin=0, ymax=1,
       xlabel near ticks,
       xlabel={data size (bytes), log scale},
       ylabel={rate (MB/s)},
       ylabel near ticks,
       xmode=log, 
			     xtick={
16,
256,
4096,
65536,
1048576,
8388608,
16777216,
268435456
},
    			  xticklabels={
16,
256,
4K,
64K,
1M,
8M,
16M,
256M},xtick scale label code/.code={},
     ]
\addplot[
		each nth point={1},
		dash dot, ACMRed,  thick, mark=*,mark options={solid, scale=0.75},
		] table[x =x, y =y]{pycrypt-one-dec.txt};
\addlegendentry{baseline (decrypt)}
       \addplot[
		each nth point={1},
		dash dot, ACMGreen,  thick, mark=square*,mark options={solid, scale=0.75},
		] table[x =x, y =y]{pycrypt-two-dec.txt};
\addlegendentry{\magic{} (decrypt)}

\addplot[
    each nth point={1},
    dash dot, ACMDarkBlue,  thick, mark=*,mark options={solid, scale=0.75},
    ] table[x =x, y =y]{pycrypt-one-enc.txt};
\addlegendentry{baseline (encrypt)}
       \addplot[
    each nth point={1},
    dash dot, ACMBlue,  thick, mark=square*,mark options={solid, scale=0.75},
    ] table[x =x, y =y]{pycrypt-two-enc.txt};
\addlegendentry{\magic{} (encrypt)}

 \addplot [name path=upper,draw=none] table[x=x,y expr=\thisrow{y}] {pycrypt-one-dec.txt};
\addplot [name path=lower,draw=none] table[x=x,y expr=\thisrow{y}-\thisrow{ey}] {pycrypt-one-dec.txt};
\addplot [fill=ACMRed!15, opacity=0.6] fill between[of=upper and lower];

 \addplot [name path=upper,draw=none] table[x=x,y expr=\thisrow{y}] {pycrypt-two-dec.txt};
\addplot [name path=lower,draw=none] table[x=x,y expr=\thisrow{y}-\thisrow{ey}] {pycrypt-two-dec.txt};
\addplot [fill=ACMGreen!15, opacity=0.6] fill between[of=upper and lower];

 \addplot [name path=upper,draw=none] table[x=x,y expr=\thisrow{y}] {pycrypt-one-enc.txt};
\addplot [name path=lower,draw=none] table[x=x,y expr=\thisrow{y}-\thisrow{ey}] {pycrypt-one-enc.txt};
\addplot [fill=ACMDarkBlue!15, opacity=0.6] fill between[of=upper and lower];

 \addplot [name path=upper,draw=none] table[x=x,y expr=\thisrow{y}] {pycrypt-two-enc.txt};
\addplot [name path=lower,draw=none] table[x=x,y expr=\thisrow{y}-\thisrow{ey}] {pycrypt-two-enc.txt};
\addplot [fill=ACMBlue!15, opacity=0.6] fill between[of=upper and lower];

	\end{axis}
\end{tikzpicture}
  \caption{\coffers with Python \textnormal{(AES cryptographic performance)}}
  \label{fig:eva:pyall}

\end{figure}
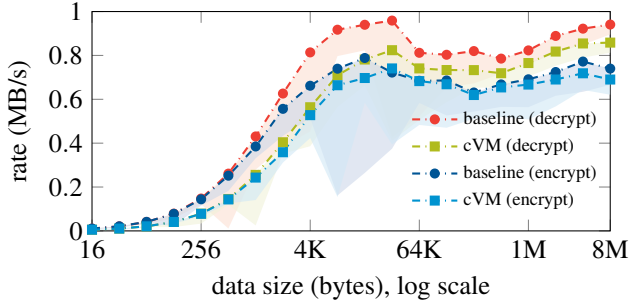

\Cref{fig:eva:pyall} shows the average throughput for encryption/decryption with different buffer sizes for \coffers, using the FPGA platform, and the baseline (non-isolated) system. Note that the low absolute numbers and variance (shown as shaded areas) are due to the platform limitations (single core), described in \cref{eval:ch}. The results in \cref{sec:eval:redis}, however, show the same trend on a platform without these limitations.

We observe that \coffers have a negligible performance impact. Throughput grows until its peak with 32\unit{KB} buffers, where the encryption/decryption rates of \coffers are only 7\% and 12\% lower than the baseline, respectively. This amounts to 0.79\unit{MB/s} and 0.96\unit{MB/s} for the baseline, and 0.74\unit{MB/s} and 0.85\unit{MB/s} for \coffers, respectively. As expected, these overheads become even smaller as the buffer sizes grow.


Our experiment shows that \acs{call} and calls into the \monitor are reasonably efficient. For reference, the mean execution time for the AES cryptographic code with a 16\unit{byte} buffer is comparable to the time for a C binding invocation in Python. At such sizes, \acs{call} invocations account for half of the overhead, which is at 97\% and 101\% for encryption and decryption, respectively, only slightly above a C binding invocation. The overhead reduces to 7\% with larger buffers.

\mypar{Conclusion} \change{\coffers{} is effective at hardening applications by isolating some of their components, such as shared libraries.} The required changes are minimal and do not change the semantics of the application interfaces, because the \acs{file} and \acs{call} primitives follow well-understood memory copy and function call semantics. Note that \acsp{stream} are constructed on top of these. The cost of this extra isolation is small, even for small buffers, and it becomes negligible as the amount of work performed between \coffers{}-enabled operations increases.






\subsection{Inter-\coffer communication}
\label{eval:microspeed}


\change[Finally, w]{W}e compare \coffers to other IPC primitives in a baseline system, and re-validate our performance results across our two platforms (FPGA and SiFive). \change{The baseline system uses two threads in a single process instead of \coffers{}; otherwise the FPGA implementation shows low TLB performance.} We measure the performance of \acsp{file} and \acsp{stream}, pipes (\texttt{PIPE}), unix sockets (\texttt{UNIX}), TCP/IP sockets (\texttt{TCP})\change[.]{\ and a combination of \texttt{mmap+memcpy+munmap} (\texttt{MAP+CPY}).}  For comparison, we also consider a raw local \texttt{memcpy} (\texttt{MEMCPY}; 4~instructions; aligned data; double-word load/store operations) as an upper performance bound. We do not evaluate \acsp{call} due to the lack of an equivalent operation in the baseline kernel.

\begin{figure}[tb]
  \centering
  \begin{subfigure}[b]{1.0\linewidth}
\centering
\begin{tikzpicture}
	\begin{axis}[
		legend columns=3,
		legend style={
    			draw=none,
		},
		legend cell align={left},
		legend style={at={(0.48,0.46)},anchor=south, font=\tiny, fill=none},
		width  = 1.0\linewidth,
		height = 5.0cm,
       xmin=2048, xmax=4194300,
       ymin=0, ymax=50,
       xlabel near ticks,
       xlabel={data size (bytes)},
       ylabel={rate (MB/s)},
       ylabel near ticks,
			     xtick={
4096,
524288,
1048576,
2097150,
2519040,
4194300
},
    			  xticklabels={
4K,
512K,
1M,
2M,
2.4M,
4M,
},xtick scale label code/.code={},
     ]

        \addplot[
		each nth point={10},
		solid, ACMDarkBlue,  very thick, mark=square*,mark options={solid, scale=0.3},
		] table[x =x, y =y2t]{micro-memcpy.txt};
\addlegendentry{MEMCPY}

          \addplot[
		each nth point={10},
		solid, ACMPurple,  very thick, mark=square*,mark options={solid, scale=0.3},
		] table[x =x, y =y2t]{micro-capfiles.txt};
\addlegendentry{\acs{file}}

        \addplot[
		each nth point={10},
		solid, ACMBlue,  very thick, mark=square*,mark options={solid, scale=0.3},
		] table[x =x, y =y]{micro-capdma-50.txt};
\addlegendentry{\acs{stream}}

     \addplot[
		each nth point={10},
		dashed, ACMGreen,  very thick, mark=*,mark options={solid, scale=0.3},
		] table[x =x, y =y]{micro-pipe.txt};
 \addlegendentry{PIPE}

     \addplot[
		each nth point={10},
		dashed, ACMRed,  very thick, mark=*,mark options={solid, scale=0.3},
		] table[x =x, y =y]{micro-local.txt};
 \addlegendentry{UNIX}

       \addplot[
		each nth point={10},
		dashed, ACMYellow,  very thick, mark=*,mark options={solid, scale=0.3},
		] table[x =x, y =y]{micro-tcp.txt};
 \addlegendentry{TCP}
 
        \addplot[
		each nth point={10},
		solid, brown,  thick, mark=square*,mark options={solid, scale=0.15},
		] table[x =x, y =y]{micro-memcpy-mmap.txt};
\addlegendentry{MAP+CPY}

	\end{axis}
\end{tikzpicture}
   \caption{FPGA CHERI}\label{fig:eva:capvstcp:fpga}
  \end{subfigure}
  \begin{subfigure}[b]{1.0\linewidth}
\centering
\begin{tikzpicture}
	\begin{axis}[
		legend columns=3,
		legend style={
    			draw=none,
		},
		legend cell align={left},
		legend style={at={(0.48,0.02)},anchor=south, font=\tiny, fill=none},
		width  = 1.0\linewidth,
		height = 5.0cm,
       xmin=2048, xmax=4194300,
       ymin=0, ymax=150,
       xlabel near ticks,
       xlabel={data size (bytes)},
       ylabel={rate (MB/s)},
       ylabel near ticks,
			     xtick={
4096,
524288,
1048576,
2097150,
2519040,
4194300
},
    			  xticklabels={
4K,
512K,
1M,
2M,
2.4M,
4M,
},xtick scale label code/.code={},
     ]

        \addplot[
		each nth point={10},
		solid, ACMDarkBlue,  very thick, mark=square*,mark options={solid, scale=0.3},
		] table[x =x, y =y2t]{sifive-micro-memcpy.txt};
\addlegendentry{MEMCPY}

          \addplot[
		each nth point={10},
		solid, ACMPurple,  very thick, mark=square*,mark options={solid, scale=0.3},
		] table[x =x, y =y2t]{sifive-micro-capfiles.txt};
\addlegendentry{\acs{file}}

        \addplot[
		each nth point={10}, filter discard warning=false,
          unbounded coords=discard,
		solid, ACMBlue,  very thick, mark=square*,mark options={solid, scale=0.3}, 
		] table[x =x, y =y2t]{sifive-micro-capdma.txt};
\addlegendentry{\acs{stream}}

     \addplot[
		each nth point={10},
		dashed, ACMGreen,  very thick, mark=*,mark options={solid, scale=0.3},
		] table[x =x, y =y2t]{sifive-micro-pipe.txt};
 \addlegendentry{PIPE}

     \addplot[
		each nth point={10},
		dashed, ACMRed,  very thick, mark=*,mark options={solid, scale=0.3},
		] table[x =x, y =y2t]{sifive-micro-local.txt};
 \addlegendentry{UNIX}

       \addplot[
		each nth point={10},
		dashed, ACMYellow,  very thick, mark=*,mark options={solid, scale=0.3},
		] table[x =x, y =y2t]{sifive-micro-tcp.txt};
 \addlegendentry{TCP}
 
        \addplot[
		each nth point={10},
		solid, brown,  thick, mark=square*,mark options={solid, scale=0.15},
		] table[x =x, y =y]{sifive-micro-memcpy-mmap.txt};
\addlegendentry{MAP+CPY}
	\end{axis}
\end{tikzpicture}
   \caption{SiFive RISC-V}\label{fig:eva:capvstcp:sifive}  \end{subfigure}
 \caption{Comparison of communication mechanisms}
 \label{fig:eva:capvstcp}
\end{figure}
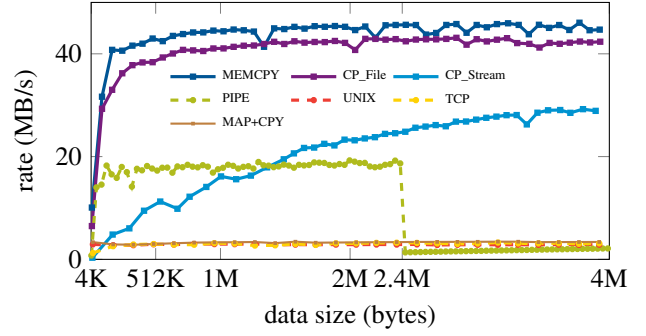
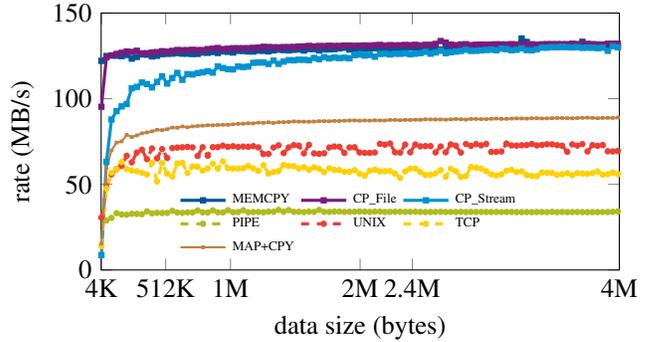

\Cref{fig:eva:capvstcp} shows the results under different buffer sizes on both the FPGA and SiFive platforms. First, the peak performance of \texttt{MEMCPY} on the FPGA platform is limited and fluctuates due to the TLB size and simple indexing function of its Flute CPU---these issues carry onto the other primitives, too.



The overhead of \acsp{file} is 6\% compared to \texttt{MEMCPY} on the FPGA platform and negligible for SiFive; it significantly outperforms all baseline IPC mechanisms. This is because we do a simple cross-\coffer \texttt{memcpy} using CHERI's \texttt{ld.cap} and \texttt{cincoffsetimm} instructions to perform the memory access and to increment the capability offset, respectively. The results also show that domain transitions via \texttt{CInvoke} are efficient, as every \acs{file} operation requires \change[two capability calls and their returns (user$\rightarrow$library OS$\rightarrow$\monitor, and back)]{one capability call and its return (user$\rightarrow$library OS, and back).}


All baseline IPC primitives have 2$\times$ overhead or more, because they perform more data copies than \texttt{MEMCPY} and \acsp{file}, closely following ideal performance. Interestingly, \acsp{stream} have worse performance on the FPGA platform, despite the lower number of copies, whereas they show performance close to \acs{file} on the SiFive platform. This is because \acsp{stream} offer an asynchronous communication primitive in which two concurrent processes time-share a single CPU core on the FPGA platform when using the \coffer API. For the same reason, all IPC primitives have lower relative performance on the FPGA platform compared to SiFive.

UNIX sockets are the closest to \acsp{stream}, because both are bi-directional, support more than two parties, and have sequenced packet modes. They exhibit only 10\% and 54\% of the performance of \acsp{stream} for 4\unit{MB} buffers on the FPGA and SiFive platforms, respectively.  \change{Here, the impact of MMU manipulation can be seen: the combination of memory copies and remapping reaches 3.4\unit{MB/s} and 89\unit{MB/s} on the FPGA and SiFive platforms, respectively. This mechanism lacks a notification primitive, and, compared to \acsp{file}, it is 15$\times$ and 1.5$\times$ slower on each platform, respectively.}

\mypar{Conclusion} For a multi-core CPU architecture with CHERI, we would expect the results to be close to those of the SiFive platform, with a minor performance decrease, similar to the difference between \texttt{memcpy} and \acsp{file} in \cref{fig:eva:capvstcp:fpga}.  This potential performance degradation is significantly smaller than the measured improvements: they range between 2$\times$ for the multi-core SiFive platform against the best baseline primitive, and 2$\times$ to an order of magnitude for the single-core FPGA platform, depending on the mechanism and buffer size.

\subsection{Deployment time}
\label{eval:teardown}

\change{We compare the deployment time of \coffers with that of Docker containers. We create a Docker image with a simple ``hello world'' program and measure the time to execute it using a \coffer and a container. For the \coffer, we use a debug-free binary with the LKL library OS and the musl standard C library ($\approx$30\unit{MB} in size) and a 10\unit{MB} application disk image. We measure two intervals, averaged over 5~runs: from the start until the output of the program, and until its termination.}

\change{On average, the Docker container requires 1.9\unit{s} to produce the output, and 2.8\unit{s} until container termination. The times for the \coffer deployment are comparable, which demonstrates their low overhead: 1.7\unit{s} and 2.6\unit{s}, respectively.}

\section{Related Work}
\label{sec:rel_work}

\mypar{Intra-process compartments} Various projects apply intra-process isolation or introduce isolation primitives. CubicleOS~\cite{sartakov2021cubicleos} isolates components of a user-level library OS using Intel MPK; unlike \coffers, it cannot readily and efficiently support legacy POSIX calls. Shreds~\cite{chen2016shreds}, Janus~\cite{janus}, Erim~\cite{vahldiek2019erim}, Hodor~\cite{hedyati2019hodor}, and Donkey~\cite{255298} use page tag-based isolation (ARM Domains, Intel MPK, or a custom RISC-V implementation) to implement protection domains and communication. In cases in which tags can be manipulated directly by user code, \eg using MPK's \texttt{wrpkru} instruction, the system requires a trusted toolchain or program verifier, unlike \coffers. Page tags also limit the number of compartments and communication buffers, as well as their granularity, which is not a problem for \coffers with capabilities.

NaCl~\cite{yee2009nacl} and WASM~\cite{haas2017wasm} face similar problems, as they require obsolete Intel segmentation and/or proof-carrying code that must be verified by a toolchain or loader. Conf\-LLVM~\cite{brahmakshatriya2019confllvm} also uses MPK to isolate code inside a process, but only supports two domains with asymmetric data exchange: trusted code can only interact with untrusted code. \coffers do not limit the number of protection domains, and inter-\coffer communication is symmetric.

LwCs~\cite{litton2016light} are an OS abstraction for intra-process protection, but they have page granularity, and switching domains comes at the cost of switching page tables. XFI~\cite{erlingsson2006xfi} provides fine-grained memory protection and control flow integrity by extending software-based fault isolation~(SFI), but SFI incurs runtime overheads and is error-prone due to its complexity.

\mypar{Compartmentalisation frameworks} \coffers{} allow the deployment of isolated shared libraries. Prior work proposes frameworks for such compartmentalization: Wedge~\cite{bittau2008wedge} identifies code parts that can be isolated; PrivTrans~\cite{brumley2004privtrans} is a source-code partitioning tool that separates trusted and untrusted components; Glamdring~\cite{lind2017glamdring} does the same for trusted execution. These approaches are orthogonal to \coffers, and they could be used to generate application components.

\mypar{Trusted execution} Intel SGX~\cite{mckeen2016intel,
  mckeen2013innovative, simon} provides \emph{enclaves} as an intra-process isolation primitive. Enclaves are part of processes and cannot be accessed by privileged software or other enclaves. Frameworks, such as Graphene-SGX~\cite{graphenesgx}, SGX-LKL~\cite{priebe2019sgx}, Panoply~\cite{shinde2017panoply}, and Spons and Shields~\cite{ssf}, deploy programs inside enclaves together with a library OS. Such designs decrease the potential impact of the untrusted OS kernel on enclaved software.

\coffers also use a library OS and share design features with these frameworks, but provide effective data sharing that cannot be implemented using enclaves. Enclaves can only share untrusted memory and cannot access each others memory, which is necessary for fast inter-\coffer communication. Since enclaves do not trust the host, they must use encryption, impacting performance~\cite{eactors}. Therefore, an interface similar to \acsp{file} cannot be implemented with enclaves.

\myparr{Library OSs} can be used to de-privilege OS kernel components or create user-level containers. $\mu$Kontainer~\cite{ukontainer} offers containers based on the LKL library OS~\cite{lkl-src}; Williams \etal\cite{10.1145/3267809.3267845} show that library OSs can be executed efficiently on top of processes instead of bare VMs; X-Containers~\cite{shen2019x} offer a cloud platform using library OSs. \coffers share similarities with user-level library OS-based containers but enhance them with strong isolation and a secure communication mechanism using capabilities.

\mypar{Machine and process isolation} As discussed in \cref{sec:background:isolation_sharing}, traditional process-based isolation has shortcomings in terms of performance and TCB size when compared to \coffers. One could envision using virtualization and Intel's \texttt{vmfunc} to strike a balance between shared TCB size and communication performance~\cite{koning2017memsentry}. Virtualization introduces well-known I/O and memory translation overheads, which are costly in a cloud stack, but are not present in \coffers.








%
%
%
%
%






\section{Conclusions}
\label{sec:concl}

\coffers are a new \change{VM-like} abstraction for cloud applications that use memory capabilities for secure isolation. \coffers include a library OS to minimize how much of the cloud environment is within the TCB. Multiple \coffers safely share an address space, allowing more efficient interaction of application components than when crossing current VM/container boundaries. Their asynchronous read/write and synchronous call interfaces allow capability-unaware, legacy code to run within \coffers.




\mypar{Acknowledgements} \change{This work was funded by the UK Government's Industrial Strategy Challenge Fund~(ISCF) under the Digital Security by Design~(DSbD) Programme (UKRI grant EP/V000365 ``CloudCAP''), and the Technology Innovation Institute~(TII) through its Secure Systems Research Center~(SSRC). It was also supported by JSPS KAKENHI grant number~18KK0310. We thank our shepherd, Ana Klimovic, and the anonymous reviewers for their helpful comments.}

\mypar{Source code availability} \change{The source code of \coffers, the Intravisor, and various application examples can be found at \url{https://github.com/lsds/intravisor}.}

\balance
\bibliographystyle{plain}
\bibliography{references.bib}

\end{document}